\documentclass{appolb}

\preprint{arXiv preprint}
\headtitle{From Spin Systems to Strongly Correlated Quantum Matter \ldots}
\headauthor{J. Spa{\l}ek}

\makeatletter
\let\Out@received\@empty
\makeatother

\usepackage[T1]{fontenc}
\usepackage[utf8]{inputenc}
\usepackage[english]{babel}

\usepackage{graphicx}
\usepackage[version=4]{mhchem}
\usepackage{amsmath,amssymb}
\usepackage{bm}
\usepackage{bbm}
\usepackage{csquotes}
\usepackage{enumitem}
\usepackage[
  backend=biber,
  style=phys,
  biblabel=brackets,
  citestyle=numeric-comp,
  sorting=none,
  minnames=4,
  maxnames=4
]{biblatex}
\usepackage[colorlinks=true,linkcolor=blue,citecolor=blue,urlcolor=blue]{hyperref}

\numberwithin{equation}{section}

\AtBeginDocument{%
\renewcommand{\figurename}{Fig.}
}

\begin{document}

\title{Strongly correlated quantum matter: t--J model, real-space pairing, spin-dependent masses, and atomicity in chemical bond and nanosystems
}
\author{Józef Spałek \thanks{Presented at {\it Concepts in Strongly Correlated Quantum Matter Conference} (CSCQM),  Kraków, Poland, 20–22 November, 2025.}%
\address{
jozef.spalek@uj.edu.pl \\ Department of Condensed Matter Theory and Nanophysics, \\ Institute of Theoretical Physics, \\ Jagiellonian University, 30-348 Kraków }
\\[3mm]
}
\maketitle
\begin{abstract}
I critically overview my research on strongly correlated fermion systems for almost five decades. It concentrated on: (i) the first derivation of what is now called the t--J model, comprising both the limit of Anderson kinetic exchange of spin--spin interaction in the Mott--Hubbard insulator and taking into account real-space pairing, subsequently applied to high-temperature superconductivity; (ii) the concept of spin-dependent heavy mass of quasiparticles in heavy-fermion systems, and (iii) the first nontrivial model of statistical thermodynamics of the Mott--Hubbard transition. Those three features, together with the specific quantum critical phenomena provide, in my view, fundamental components of the theory of strongly correlated fermions established in the 1960s.
Some related questions such as introduction of atomicity in the chemical bonding (iv), and specific properties of correlated nanosystems within the rigorous EDABI (v) approach are also briefly elaborated at the end.
\end{abstract}
  
\section{Introduction}
The concept of correlated electron (fermion) systems started with the work of Mott \cite{Mott1997n} who pointed out that the electron--electron repulsive Coulomb interaction may be responsible for the breakdown of single-particle (band) theory of systems such as the transition metal oxides CoO or NiO. In effect, the Hund's rule type of $3d$-ion configuration (i.e., that with maximal spin of singly occupied $3d$ shells) may arise and is caused by the strong intraatomic repulsion $U$, supplemented with a weaker interorbital ferromagnetic exchange 
$ \sim J^H \sim U/4$ (U is the magnitude of intraorbital Coulomb repulsion
and is of the order 5--10 eV). 
These qualitative concepts have been formalized and developed in the 1960s within the second quantization language by Anderson \cite{Anderson1959, Anderson1963}, Hubbard \cite{Hubbard1964}, Kanamori \cite{Kanamori1963}, and Gutzwiller \cite{Gutzwiller1965}. The new approach required application of the methods going beyond the standard perturbation theory in quantum mechanics, since the magnitude of the intraatomic (Hubbard) interaction $U$ was comparable or even remarkably higher than the bare Fermi energy $\epsilon_F \sim 5-10 eV$ of electrons composing a collective liquid in those systems. In this situation, the criteria (Mott and Hubbard) localization have been invoked for  electron persistence in atomic states even when they compose a solid. Such
criteria have been first established for correlated-electron system in the ground state.
Those global conditions have been complemented with the result of Brinkman and Rice \cite{Brinkman1970}, that the effective mass of the correlated carriers diverges at that localization threshold and is the source of such a continuous phase transition in paramagnetic phase, since associated with this mass effective Pauli susceptibility also diverges.
In this manner, the study of purely electronic properties of those systems with narrow bands 
have merged with those of quantum phase transitions, i.e., those with specific quantum fluctuations, arising from competing dynamical processes even at temperature $T=0$. To end this historical note, our first successful description of the metal-insulator (Mott--Hubbard) transition at temperature $T>0$ \cite{Spalek1987,Spaek1989} was briefly reviewed in \cite{Spalek1990}. The same physics may be applied to other topics concerning fermion systems such as liquid ${}^{3}\mathrm{He}$ \cite{Wysokinski2014} or cold atomic systems \cite{Joerdens2008}. 

The present overview complements the three published earlier similar accounts \cite{Spalek2007,Spalek2012,Spalek2023} in that the present one summarizes our work from physical point of view, as well as discusses problem of some new subsequent work being developed in our group in relation to high-temperature (high-${\rm T_c}$) superconductivity.

Namely, the topics comprise: The spin--direction dependent effective masses in magnetically polarized systems, and the kinetic exchange interaction that leads to the specific real-space pairing, resulting in an unconventional (high-temperature) superconductivity. It is the author's view that these phenomena, together with specific type of quantum critical behavior in those systems, determine the essence of the strong-correlation physics paradigm. The separate part of the paradigm is devoted first of all largely computational methods based on extensions of (single-particle) DFT method, i.e., DFT-DMFT, DFT + U, specific quantum Monte-Carlo methods, as well as renormalization group methods \cite{Jones2015,2013}. Those methods will not be discussed here, apart from our own method of EDABI ({\bf E}xact {\bf D}iagonalization {\bf A}b {\bf I}nitio), to which the author have contributed originally, particularly as applied it  to the nanophysical systems \cite{Spalek2020} and to the theory of chemical bonding. In what follows we discuss these topics separately.

An additional remark is in order at this point. The strongly correlated quantum electronic materials are customarily divided into two classes: {\it the narrow band materials}, in which the 
electronic narrow band(s) play a dominant role (the Mott--Hubbard materials) and those in which the strongly correlated electrons are hybridized in a nontrivial way with uncorrelated valence-band electrons; these are called {\it the heavy-fermion materials}. The physics of those two systems is related, as will be discussed briefly later.

\section{Kinetic exchange: t--J Hamiltonian and its extensions}
\subsection{Direct versus kinetic exchange}
It is important to state at the beginning that the exchange interaction is a universal feature of fermions, since it  follows directly from the antisymmetric character of many-particle wave function for indistinguishable fermions. One can note that the spin indistinguishability is sufficient for the exchange interaction to appear explicitly even if the fermions are orbitally distinguishable, as is the case of the Fermi contact proton--electron spin-dependent interaction in atoms. Originally, for the author an enlightening discussion of that fundamental property of fermions turned out to be the Dirac formulation \cite{Dirac2010}, particularly reformulated in the second-quantization formalism by V.A. Fock \cite{Fock1957book}. A schematic representation of the situation discussed here is shown in Fig. 1. 

\begin{figure}[htb]
\centering
\includegraphics[width=0.9\textwidth]{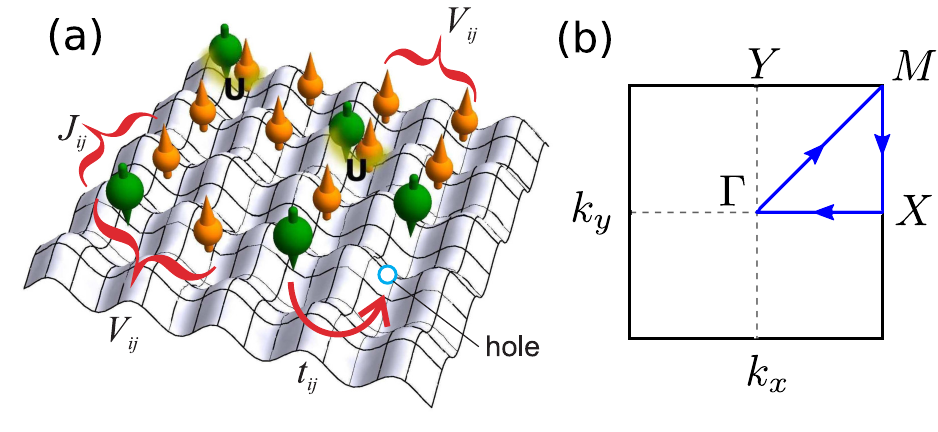}
\caption{Left: Visualization of a quantum Fermi liquid as an effective lattice gas of 
spin-$\frac{1}{2}$ particles. Every local minimum can accommodate up to two fermions 
with opposite spins. The Hubbard U intraatomic, as well as those between the 
nearest neighbors $\langle i j \rangle$ are marked (see main text). The number of
particles is smaller than the number of atomic sites. Right: the first Brillouin zone 
for square lattice with marked high-symmetry points \cite{Wysokinski2014}.}
\label{Fig1}
\end{figure}

Strictly speaking, the Heisenberg--Dirac definition of exchange interaction is strictly valid to the fermions with spin $S = 1/2$ in \textit{orbitally fixed} (atomic) states. This is the reason for calling it a direct exchange. Its exchange coupling constant (integral) is determined 
solely by the Coulomb repulsive interaction and the explicit form of the orbital wave function. The Fock formulation based in the second-quantization language allows for its generalization to the situation when the particles are also itinerant, i.e., change their orbital state due to quantum-mechanical hopping between those states. This is the situation when the kinetic exchange comes into play. In that situation particles are indistinguishable with respect to both orbital and spin degrees of freedom. Then, the kinetic exchange interaction should reduce to the former Heisenberg--Dirac form in the situation when the orbital degrees of freedom become frozen. This is exactly the situation that happens when the correlated itinerant fermionic liquid undergoes a transition to the Mott--Hubbard localized state. In the latter situation the orbital dynamics of particles is reduced to the \textit{virtual hoppings} only between the different orbital states. This was also exactly the original point of departure of our construction \cite{Spalek2007modifiedacta} to the t--J model, where the reference point contained the kinetic exchange considerations of Anderson for Mott insulators \cite{Anderson1963}.

\subsection{Physical essence of the strong-correlation limit $|t_{ij}| \ll  U$}
Discussion of the t--J model starts usually from the Hubbard model, although a similar formalism has been extended also to the Anderson-lattice model of hybridized correlated and uncorrelated carriers (see below). The Hubbard model for spin $S = 1/2$ particles has the form
\begin{equation} \label{eq:ham}
    \hat{\mathcal{H}} = \epsilon_{at} \sum_{i \sigma} \hat{n}_{i \sigma} + \sum_{ i j \sigma} t_{ij} \hat{a}^\dagger_{i \sigma} \hat{a}_{j \sigma} + U \sum_i \hat{n}_{i \uparrow} \hat{n}_{i \downarrow}.
\end{equation}
In this expression the first term represents the reference point (atomic) energy $\epsilon_{\rm at}$, the second (for $i\neq j$) provides the single-particle part of hopping fermions $j \to i$ with spin quantum number $\sigma = \pm 1 = 2 s^z $ and hopping amplitude $t_{ij} < 0$. The last provides the repulsive ($U>0$) Coulomb interaction if two electrons (fermions) with opposite spins meet 
the same Wannier orbital. The $\epsilon_{at}$ can be taken as reference point and put $\epsilon_{at} = 0$, since total number of particles is conserved in a periodic solid, i.e., $\sum_{i \sigma} \hat{n}_{i \sigma} = N_e = \rm{const}$, where $N_e$ is the total number of particles in the system of N sites. The hopping $t_{ij}$ and the Hubbard parameter $U$ are defined as $t_{ij} \equiv \langle w_i | \mathcal{H}_i | w_i \rangle$ and $U \equiv \langle w_i w_i | \mathcal{H}_2 | w_i w_i \rangle$, respectively, where $w_i \equiv w(\bf{r} - \bf{R}_i)$ represents the orbital part of single particle (Wannier) wave function for the corresponding Hamiltonian $\mathcal{H}_1$ for that particle. As a rule, the set $\{ w_i (\bf{r})\}$ is regarded as orthogonal and normalized, $\langle w_i |w_j \rangle = \delta_{ij}$. 

A detailed discussion starts with the definition of the strong-correlation limit. For that purpose one defines the bare band width in the form $W \equiv 2 \sum_{j(i)} |t_{ij}|$ of uncorrelated particles, where the summation is over all the neighboring sites $j$ of given central site $i$.  The strongly correlated limit is defined then as that with $W \ll U - K$, with $K$ being the amplitude of nearest-neighbor Coulomb repulsive interaction, $K\equiv \langle w_i w_j | H_2 | w_i w_j \rangle$. In practice, 
it is sufficient to assume that $W$ is substantially smaller then $(U - K)$, as the most important thing in actual canonical perturbation expansion we should have is $|t_{ij}| \ll (U-K)$. 

The fundamental formal starting point is to decompose the dynamics of correlated itinerant particles into two parts, i.e., that in either low- and high-energy Fock subspaces. This amounts to the decomposition of the creation (and annihilation) operators in the following way
\begin{equation}
    \hat{a}^\dagger_{i \sigma} \equiv \hat{a}^\dagger_{i \sigma} (1 - \hat{n}_{i \overline{\sigma}} + \hat{n}_{i \overline{\sigma}}) \equiv \hat{a}^\dagger_{i \sigma} (1 - \hat{n}_{i \overline{\sigma}})+ \hat{a}^\dagger_{i \sigma} \hat{n}_{i \overline{\sigma}},
\end{equation}
and in the same manner, the particle number operator
\begin{equation}
    \hat{n}_{i \sigma} \equiv \hat{n}_{i \sigma} (1 - \hat{n}_{i \overline{\sigma}}) + \hat{n}_{i \sigma} \hat{n}_{i \overline{\sigma}}.
\end{equation}
In other words, we have divided the local operators into two parts: the first with no site (local) double occupancy with the opposite spins involved and that having them. Such a decomposition is in direct analogy to the decomposition of fermionic states into the Hubbard subbands \cite{Hubbard1964}, here transposed to the local (site) language. Equivalently, one can define the global projection operators $P_1$ and $P_2$ through the identity 
\begin{equation}
    P \equiv \prod_{i \sigma} \left[ \hat{n}_{i \sigma} ( 1 - \hat{n}_{i \uparrow} \hat{n}_{i \downarrow}) + x \hat{n}_{i \uparrow} \hat{n}_{i \downarrow} \right] \equiv \sum_{l = 0} P_l x^l \equiv P_1 + P_2,
\end{equation}
where $P_1$ corresponds to the states with zero- and single-site occupancies ($l=0$) and $P_2$ to those with at least one ($l = 1$) double, as well as with all those higher double occupancies ($l > 1$). Note that in the interesting us situation with $x \equiv 1$ we have $P_1 + P_2 \equiv \mathbbm{1}$ and  $P_1 \cdot P_2 = \emptyset$, so $\hat{P}_1$ and $\hat{P}_2$ projections decompose the N-site Fock space into two orthogonal subspaces. Also, $P_1 = P_1^\dagger$ and $P_2 = \hat{P}_2^\dagger$ so that the projected Hamiltonian components $P_i \mathcal{H} P_j$ preserve their hermicity.

The Hamiltonian (\ref{eq:ham}) when decomposed into the two above parts contains, nonetheless, the processes mixing the subspaces, i.e.,
\begin{equation}\label{eq:ham1}
    \hat{\mathcal{H}} \equiv ( P_1 + P_2) \hat{\mathcal{H}} (P_1 + P_2) = P_1 \hat{\mathcal{H}} P_1 + P_2 \hat{\mathcal{H}} P_2 + P_1 \hat{\mathcal{H}} P_2 + P_2 \hat{\mathcal{H}} P_1.
\end{equation}
At this point a physical remark is in place. First, the Hamiltonian starts from the atomic picture as the model is defined in terms of Wannier (orthogonalized atomic) states that lead to the model in the tight-binding real-space language, since we usually limit the hopping integrals to nearest $\langle i,j \rangle$ or next-nearest neighbors. Likewise, the interaction integrals are limited to the intraatomic (Hubbard model case) or the nearest-neighbor sites (extended Hubbard or t--J, and related models). Such formulation differs essentially from the standard approach when one starts from the concept of electron gas or even Fermi liquid, where the complementary, in the quantum mechanical sense momentum representation is assumed as valid throughout analysis, independently 
of the interaction strength. Obviously, the approach starting from atomic (site) representation must reproduce the principal physical properties of the Fermi liquid theory in the low-interaction (low-correlation) limit, at least that in the tight-binding approximation.

\subsection{Physical analysis of t--J model derivation and context}
To derive the form of the effective (t--J and the like) Hamiltonian formally within the so-called canonical perturbation expansion \cite{Spalek1977,Chao1977,Chao1978} we have to treat the last two terms in (\ref{eq:ham1}) as a perturbation. Note that the perturbation part  $P_1 \mathcal{H} P_2 + P_2 \mathcal{H} P_1$ contains both terms
$\sim t_{ij}$  and  those $\sim U$ terms, whereas $P_1 \mathcal{H} P_1$ contains only processes $\sim t_{ij}$. So, the perturbation expansion is justified physically by a large energy distance between the states of $P_1 \hat{\mathcal{H}} P_1$ composing the initial state with respect to those composing  $P_2 \hat{H} P_2$ and regarded as intermediate states. In other words the small parameter $t_{ij}$ with respect to $U$ is not a direct perturbation parameter. In the literature the latter view $|t_{ij}| \ll U$ is commonly stressed, not quite precisely. We return
to this point later on. 

From the formal point of view, the projected components of (\ref{eq:ham1}) have the explicit forms:
\begin{align}
    P_1 \hat{H} P_1 & \equiv P_1 \sum_{i j \sigma} t_{ij} \hat{a}^\dagger_{i \sigma} (1 - \hat{n}_{i \overline{\sigma}}) \hat{a}_{j \sigma} (1 - \hat{n}_{j \overline{\sigma}}) P_1, \\ 
    P_2 \hat{H} P_2 & \equiv P_2 \left\{ \sum_{i j \sigma} t_{ij} \hat{a}^\dagger_{i \sigma} \hat{n}_{i \overline{\sigma}} \hat{a}_{j \sigma} \hat{n}_{j \overline{\sigma}} + U \sum_i \hat{n}_{i \uparrow} \hat{n}_{i \downarrow}\right\} P_2, \\ 
    P_1 \hat{H} P_2 & \equiv (P_2 \hat{H} P_1)^\dagger \equiv P_1 \left\{ \sum_{i j \sigma} t_{ij} \hat{a}^\dagger_{i \sigma} (1 - \hat{n}_{i \overline{\sigma}}) \hat{a}_{j \sigma} \hat{n}_{j \overline{\sigma}} \right\} P_2.
\end{align}
Note that the formal presence of the global projection $P_i$ in the expressions guarantee that the sites which do not participate in given local hopping $i \leftrightarrow j$ term belong to a proper (lower energy) subspace with number $l=0$ ($P_1$) and $l>0$ ($P_2$) this formal requirement is ignored in our original derivation \cite{Spalek1977,Chao1977} since we have been interested only in the second order results from the purely formal side.

A detailed derivation of the effective Hamiltonian $\hat{H}$ contains canonical perturbation expansion in which the part ($P_1 \mathcal{H} P_2 + P_2 \mathcal{H} P_1$) part is removed by canonical perturbation expansion and replaced by the low-energy ($P_1 \tilde{\mathcal{H}} P_1$) and high-energy parts ($P_2 \tilde{\mathcal{H}} P_2$) in \textit{all orders}, as guaranteed by the successive projections. The details have been carried out repeatedly before \cite{Spalek2007modifiedacta, Chao1978}, so we discuss here the results in the first nontrivial (second) order, $P_1 \hat{\mathcal{H}} P_1$. The effective Hamiltonian $P_1 \tilde{\mathcal{H}} P_1$ has the following explicit form
\begin{align} \label{eq:ham2}
    P_1 \tilde{H} P_1 & \quad = P_1 \Bigr\{ \sum_{i j \sigma} t_{ij} \hat{a}^\dagger_{i \sigma} (1 - \hat{n}_{i \overline{\sigma}}) \hat{a}_{j \sigma} (1 - \hat{n}_{j \overline{\sigma}})  \nonumber \\
    & \quad + \sum_{ij} \frac{2 t^2_{ij}}{U} \left[ \mathbf{S}_i \cdot \mathbf{S}_j - \frac{1}{4} \sum_{\sigma \sigma'} \hat{n}_{i \sigma} (1 - \hat{n}_{i \overline{\sigma}}) \hat{n}_{j \sigma} (1 - \hat{n}_{j \overline{\sigma}}) \right]   \nonumber \\
    & \quad + \sum_{ijk \sigma \sigma'} \frac{t_{ij} t_{jk}}{ U} \hat{a}^\dagger_{j \sigma'} (1 - \hat{n}_{j \overline{\sigma}}) \hat{a}_{k\sigma'} \hat{n}_{k \overline{\sigma}} \hat{a}^\dagger_{k \sigma} \hat{n}_{k\overline{\sigma}} a_{i\sigma} (1 - \hat{n}_{i \overline{\sigma}})\Bigr\} P_1.
\end{align}
In this Hamiltonian the first term represents the projected hopping between the simple occupied and empty sites, the second-order virtual hopping forth and back between the sites $\langle i, j \rangle$, and the third represents the three-site hopping $i\to j \to k$. Those processes are schematically represented 
in Fig. 2. Note that the first and the last diagrams are only present in a partially filled band case with $n_i \equiv \langle \hat{n}_{i \sigma} + \hat{n}_{i \overline{\sigma}} \rangle < 1$. Actually, if the number of particles is strictly preserved on each site, i.e., when $\hat{n}_i = \hat{n}_{i\uparrow} + \hat{n}_{i \downarrow} = 1$, the Hamiltonian reduces to the Anderson kinetic-exchange Hamiltonian
\begin{equation}
    P_1 \hat{\mathcal{H}} P_1 = \sum_{ij} \frac{2 t^2_{ij}}{U} (\mathbf{S}_i \mathbf{S}_j - \frac{1}{4}),
\end{equation}
describing the antiferromagnetic exchange in Mott insulator, corresponding to the situation with one electron per atom, $\hat{n}_i \equiv 1$ (half-filled narrow band case). In the situation with orbitally degenerate orbitals the situation is the same although it must involve also the Hund's rule coupling. Also, the derivation of  (\ref{eq:ham2}) can be extended to the case when we include all pair interactions in the starting Hamiltonian and then the result corresponding to (\ref{eq:ham2}) is discussed in detail in \cite{Spalek1980} for orbitally degenerate systems as well.

\begin{figure}[htb]
\centering
\includegraphics[width=0.9\textwidth]{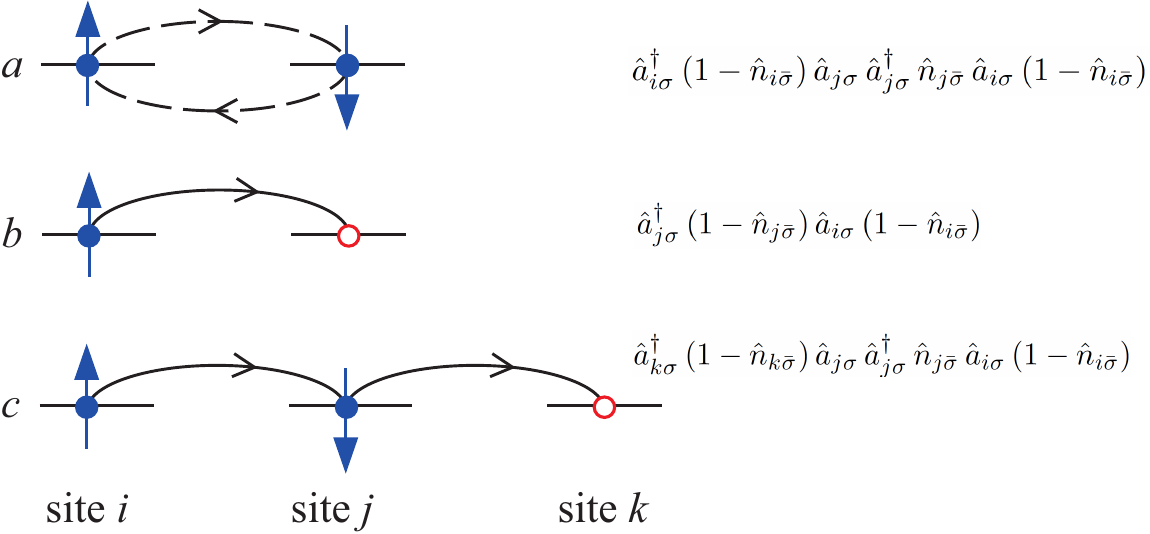}
\caption{Various neighboring hopping processes: a virtual hopping, b-real hopping, and c-three-site hopping in the second order. In the limit of Mott insulator only the first term (a)
survives and leads to the pure kinetic-exchange interaction form.}
\label{Fig2}
\end{figure}

In the expressions (\ref{eq:ham2}) the spin operators defined in the fermion representation, i.e.,
\begin{align}
    \hat{\mathbf{S}}_i \equiv & (\hat{S}^+_i,\hat{S}^-_i,\hat{S}^z_i) \equiv (\hat{a}^\dagger_{i\uparrow}\hat{a}_{i\downarrow}, \hat{a}^\dagger_{i\downarrow}\hat{a}_{i\uparrow},\frac{1}{2} (\hat{n}_{i\uparrow} - \hat{n}_{i\downarrow})) \nonumber \\ \equiv &  \left(\hat{b}^\dagger_{i\uparrow}\hat{b}_{i\downarrow}, \hat{b}^\dagger_{i\downarrow}\hat{b}_{i\uparrow},\frac{1}{2} (\hat{\nu}_{i\uparrow} - \hat{\nu}_{i\downarrow})\right).
\end{align}
This definition is the same in both the original fermion operators (as above), as well as in terms of projected fermion operators $\hat{b}^\dagger_{i\sigma} \equiv \hat{a}^\dagger_{i\sigma} (1 - \hat{n}_{i \overline{\sigma}})$, $\hat{b}_{i\sigma} \equiv \hat{a}_{i\sigma} (1 - \hat{n}_{i \overline{\sigma}})$, and the corresponding particle number operators $\hat{n}_{i \sigma} \to \hat{\nu}_{i\sigma} \equiv \hat{n}_{i \sigma} (1 - \hat{n}_{i \overline{\sigma}})$. In the strict limit of Mott insulator $\hat{n}_{i \uparrow} + \hat{n}_{i \downarrow} = 1$ they 
reduce to the Pauli operators $\hat{\mathbf{S}}_i = \frac{1}{2} \hat{\bm{\tau}}$, where $\hat{\bm{\tau}}$ denotes the respective Pauli matrices. 
An indirect proof of this reduction is provided by calculation $\langle \mathbf{S}_i^2 \rangle$ in the ground state \cite{Spalek1990}. Within the mean-field (Gutzwiller) approximation we obtain that for $n=1$ that $\langle \mathbf{S}_i^2 \rangle = \frac{3}{4} (1 - 2 d^2)$, where $d^2 \equiv \langle n_{i\uparrow} n_{i \downarrow} \rangle$. 
In the Mott limit $d^2  \to 0$ and then $\langle \mathbf{S}_i^2 \rangle = \frac{3}{4} = \frac{1}{2} (\frac{1}{2} + 1)$, i.e., the localized moment value is fully recovered.

\subsection{Projected real-space pairing operators: Quantum spin liquid or Cooper pairs?}
From the foregoing discussion it follows that the Hamiltonian (\ref{eq:ham2}) describes the dynamics of electrons at least at and close to the Mott-insulator limit, i.e., for $n \equiv \langle n_{i \uparrow} + n_{i \downarrow} \rangle \leq 1$, in the projected lowest-energy Fock subspace (corresponding to the lower Hubbard subband). The proper single-particle fermion operators are then $\hat{b}_{i \sigma}^\dagger \equiv \hat{a}^\dagger_{i \sigma} (1 - \hat{n}_{i \overline{\sigma}})$ and its Hermitian conjugate $\hat{b}_{i \sigma}$. It was surprising to the author to discover that a relatively complex expression of the effective Hamiltonian (\ref{eq:ham2}) can be brought to the simpler and formally closed form by introducing also the projected spin-singlet pairing operators in the real-space \cite{Spalek1988}
\begin{align}
    \begin{cases}
        \hat{b}^\dagger_{i j} \equiv \frac{1}{\sqrt{2}} (\hat{b}^\dagger_{i \uparrow} \hat{b}^\dagger_{j\downarrow} - \hat{b}^\dagger_{i \downarrow} \hat{b}^\dagger_{j\uparrow}), \\
        \hat{b}_{i j} \equiv (\hat{b}^\dagger_{i j})^\dagger  = \frac{1}{\sqrt{2}} (\hat{b}_{i \uparrow} \hat{b}_{j\downarrow} - \hat{b}_{i \downarrow} \hat{b}_{j\uparrow})
    \end{cases}
\end{align}
Note that $\hat{b}_{ii} \equiv 0$, so the pairing vanishes identically on the same site. In effect, (\ref{eq:ham2}) takes the closed form 
\begin{equation}
    P_1 \tilde{\mathcal{H}} P_1 = P_1 \Bigr\{ \sum_{ij \sigma}  t_{ij} \hat{b}^\dagger_{i \sigma} \hat{b}_{j\sigma} - \sum_{i,j,k} \frac{2 t_{ij} t_{jk}}{U} \hat{b}^\dagger_{i j} \hat{b}_{kj} \Bigr\} P_1,
\end{equation}
where now the second-term contains the two previous interaction terms for $k=i$ and $k\neq i$, respectively. From this expression we see explicitly that the first term describes, as before, the restricted single-particle hopping, whereas the second contains both the binding part ($\sim \hat{b}^\dagger_{ij} \hat{b}_{ij}$) of the local pairs in the spin-singlet configuration, as well as the local pair hopping for $k\neq i$ ("waltz dancing" of each pair). As a result we have a clear competition between the single-particle hopping and bound local singlet pair rotative dancing. In our model based on (\ref{eq:ham2}), with no admixture of site double occupancies; their presence is necessary if we have $n>1$, but then one has to bring into the projected part $P_2 \tilde{\mathcal{H}} P_2$ picture. 
Nonetheless, even when we limit ourselves to the model based on the Hamiltonian (\ref{eq:ham2}) only, we can see clearly that the three different types
of ordering are possible. First of them is the antiferromagnetic state, 
which should appear for $n=1$ and then one can expect it can also survive close to that limit
for $n \lesssim 1$ (see also below). The second expected state is the \textit{superconducting} state, which can be characterized by $\langle \hat{b}^\dagger_{ij} \rangle \neq 0$ as the local pairing amplitude. But then, it can be also identified as the \textit{quantum-spin-liquid} state with incoherent local singlet-pairs dynamics by consecutive pair-spin flips. Such a state is customarily called the \textit{resonating valence bond state} (RVB). Obviously, the paired superconducting state can be properly characterized by the nonzero off-diagonal long-order correlation function \cite{Yang1962} $\langle \hat{b}^\dagger_{ij} \hat{b}^\dagger_{mn} \rangle$ in the limit $| \mathbf{R}_{ij} - \mathbf{R}_{m n}| \to \infty$, but that is not easy to
determine for an extended system. In that category of thinking, the RVB state can be distinguished from superconducting spin-singlet pairing by determining the diagonal-order correlation function $\langle \hat{b}^\dagger_{i j} \hat{b}_{m n} \rangle$ as a function
of relative distance $ | \mathbf{R}_{i j} - \mathbf{R}_{m n}| $. In  the mean-field approximation
when the two-particle correlations are factorized e.g. by a decompositions $\langle \hat{b}^\dagger \hat{b}^\dagger \rangle \approx \langle \hat{b}^\dagger\rangle \langle \hat{b}^\dagger \rangle$ or $\langle \hat{b}^\dagger \hat{b} \rangle \approx \langle \hat{b}^\dagger\rangle \langle \hat{b} \rangle$, the distinction between local pairing and the RVB cannot be clearly made, particularly in two dimensions. The possible distinction 
may be then made probably by the phase difference $\varphi_{ij} - \varphi_{mn}$
of the respective correlation functions.

Next, we briefly overview an alternative way in the next section in the form of systematic \textbf{D}iagramatic \textbf{E}xpansion of Variational (\textbf{G}utzwiller) \textbf{W}ave \textbf{F}unction (DE-GWF). But first we have to change our view on the restriction on local double occupancy. This is 
needed to be able to treat both hole- and electron-doped systems on the same footing.

\subsection{Extension: t--J--U--(V) model}
Working directly with the projected fermion operators $\hat{b}_{i\sigma}$ (and $\hat{b}^\dagger_{i \sigma}$) is cumbersome. The essence of the problem is that those operators have non-fermion anticommutation relations, namely
\begin{equation}
    \Bigr\{ \hat{b}_{i \sigma},\hat{b}^\dagger_{i \sigma} \Bigr\} = \delta_{ij} \left[ (1 - \hat{n}_{i \overline{\sigma}}) \delta_{\sigma\sigma'} + \hat{S}_i^{\overline{\sigma}} (1 - \delta_{\sigma \sigma'} )\right],
\end{equation}
which are difficult to handle, In effect, we have proposed to extend our
approach in the following manner. Namely, we combine t--J model with the original Hubbard model by constructing an effective t--J--U model with the starting Hamiltonian of the form \cite{Spalek2017}
\begin{equation} \label{eq:ham3}
    \hat{H} = \sum_{ij\sigma}\!\raisebox{1.4ex}{$\prime$} t_{ij} \hat{a}_{i \sigma}^\dagger \hat{a}_{j \sigma} + \sum_{ij}\!\raisebox{1.4ex}{$\prime$} J_{ij} \left( \hat{\mathbf{S}}_i \cdot \hat{\mathbf{S}}_j  - \frac{1}{4} \hat{n}_i \hat{n}_j \right) + U \sum_i \hat{n}_{i \uparrow} \hat{n}_{i \downarrow}.
\end{equation}
Formally, this Hamiltonian encompasses the cases of the Hubbard model limit 
for $J_{ij}\equiv 0$ and that of t--J model for $U \to \infty$. It has the advantage that the cumbersome projection onto the $P_1$ space with is not necessary, as we deal with ordinary fermion $a_{i \sigma}^\dagger$ and $a_{i\sigma}$ operators. 
However, then a subtle point remains to explain that $J_{ij}$ and $U$ terms appear at the same time. Our explanation of the last fact is that we treat the kinetic exchange term as appears then as a perturbation, also  by taking into account the contribution of superexchange via other states, primarily of $2p$ character (i.e., via oxygen). In other words, we regard the Hubbard model as intrinsic, with physical value of $U \sim 8-10~eV$ for $3d$ electrons located on copper, with small $d-d$ direct contribution via $ 3d  \leftrightarrow 3d$ virtual hopping, particularly 
if we are to take the realistic U value, but with the dominant $d_i \leftrightarrow 2 p \leftrightarrow 2d_j$ superexchange contribution. 
Furthermore, additional advantage of the model (\ref{eq:ham3}) is that we can analyze the system in the full band filling range $0 \leq n \leq 2$, comprising both the hole- and electron-doped regimes for the high-$\rm{T_c}$ 
cuprates \cite{Zegrodnik2017,Dey2026}.

One remark is in place here. As the phase diagram of the cuprates is quite involved and encompasses antiferromagnetic insulating, high-temperature
superconducting, and charge-density-wave phases, it is
important to additionally include direct Coulomb intersite interaction, i.e., the term $\sum_{ij}' V_{ij} \hat{n}_i \hat{n}_j$ \cite{Dey2026}.
Then, the $V_{ij}$ term contains also part of the kinetic exchange, $- \sum_{i j}' J_{ij} \hat{n}_{i} \hat{n}_j$. All these factors will be briefly touched upon in the next section, where we discuss our selected results.
\subsection{DE-GWF variational approach and results: A summary}
\subsubsection{\centering Equilibrium properties}
The second crucial component of the proposed microscopic model
is its solution and comparison with experimental results for high-${\rm T_c}$ cuprates. As this overview is concerned with the results of the author's group, we concentrate here on brief characterization of our systematic variational approach, starting from a modified Gutzwiller approach (GA) in the form of statistically consistent Gutzwiller approximation (SGA) and then summarizing the higher-order results within diagrammatic expansion of the (SGA-modified) Gutzwiller approach (DE-GWF);
for details see \cite{Spalek2017, Zegrodnik2017, Dey2026, Spalek2022}.

First, let us discuss the question why the original Gutzwiller (GA) approach had to be modified. Our early simple analysis \cite{Kaczmarczyk1008} of the plain Gutzwiller approximation has resulted in the dilemma that the approach based on solving the resulting self-consistent equations for physical quantities leads in some cases to the final results that differ from those obtained from the direct variational solution. We stress that agreement between the two approaches is the necessary condition for quasiparticles to be defined properly (we called this the Bogoliubov consistency condition). To achieve that, we were forced to introduce the corresponding constraints to ensure the agreement between the two results of the two methods and thus achieve the statistical-mechanical consistency. Hence, the name of our modified mean-field (GA) approach as statistically consistent variational Gutzwiller approximation (SGA). It should be noted that the SGA method leads formally  to the results of the same type as those obtained in the slave-boson approach in the saddle-point approximation \cite{Kotliar1986,SpalekWojcik1995n}. However, in our approach only physical fields appear, i.e., without ghost Bose fields and therefore, spurious Bose condensation of them is absent, as should be \cite{SpalekWojcik1995n}.

The method of formal analysis of SGA and its systematic diagrammatic expansion of variational wave function (DE-GWF) is presented in detail elsewhere \cite{Spalek2022}. Here we summarize only the exemplary results for 
the case of high temperature superconducting cuprates and compare some of them directly with experiment. In Fig. 3 (a) and (b) we present the phase diagram obtained
within extended t--J model (a) as compared to the experimental results (b). The results include the appearance of the pair density wave (PDW) phase, as labeled by the contribution $\delta \Delta^d_{\rm PDW}$. 
Note that the appearance of PDW suppresses partially the pure d-wave characterized by $\overline{\Delta}^d$. The remaining (minor) alternative phases do not reproduce the overall experimental behavior of the superconducting gap as a function of the doping $\delta \equiv 1 - n$. The
qualitative agreement, apart from the absence of the pseudogap (dotted line
in (b)) is good, though the upper critical concentration for d-wave superconductivity disappearance, is too high in (a).

\begin{figure}[htb]
\centering
\includegraphics[width=0.9\textwidth]{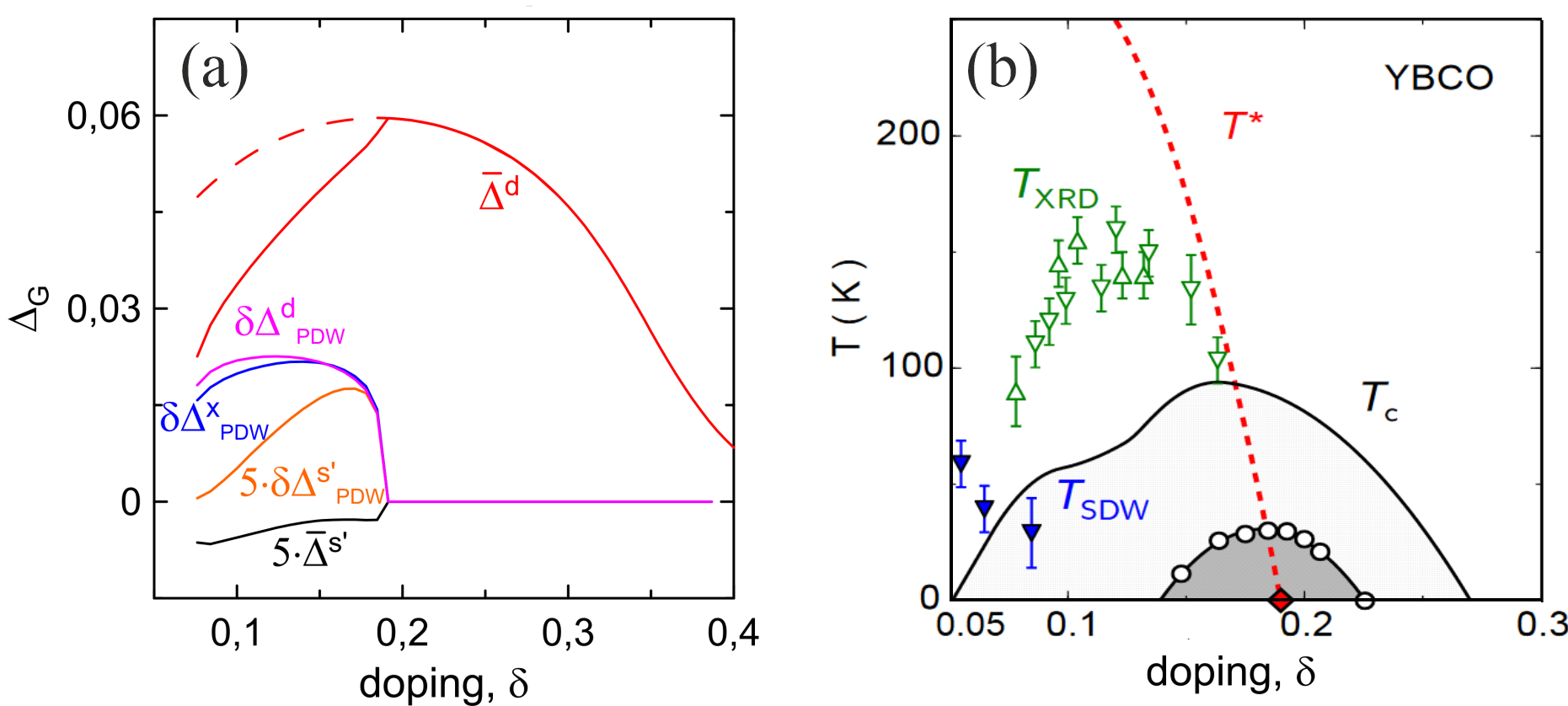}
\caption{The phase diagram comprising charge-density-wave (CDW) states: (a) theory \cite{Zegrodnik2018}; (b) experiment \cite{Bialo2020}. The onset of the pair density wave (PDW) has the largest impact on the left part of the diagram  (a) and induces a small s-wave pairing component in the dominant d-wave SC state, which survives as a pure state at and above the optimal doping. For details see \cite{Zegrodnik2018,Bialo2020}.}
\label{Fig3}
\end{figure}

The extended t--J--U--(V) model allows for characterization of the paired superconducting phases for both hole- and electron-doped regimes for the cuprates, as shown in Fig. 4 (a) and (b) for both $\delta \ge 0$ (hole doped) and $\delta \le 0$, respectively. The lower part (b) includes the situation when the so-called pair-hopping term is also included \cite{Zegrodnik2017}.
Note, as we have mentioned earlier, the extension of our canonical t--J model
into t--J--U--V version and related form allows for tracing the phase diagram both in the hole $\delta = 1 - n \ge 0$ and electron $\delta = 1 - n \le 0$ regimes, with the vanishing pairing in the Mott-insulating limit ($\delta =0$). Furthermore, in the upper part (a) we have the BCS-like and non-BCS-like regions; the former being characterized as the one in which the kinetic energy of the carriers is lowered during the transition to the d-wave paired state, whereas in the BCS-like regime the potential (pair-binding) energy gain takes place, as evidenced explicitly in the  BCS theory by appearance of the superconducting gap, lowering the energy of the occupied states by amount $\simeq \Delta$ in the latter case. 

\begin{figure}[htb]
\centering
\includegraphics[width=0.9\textwidth]{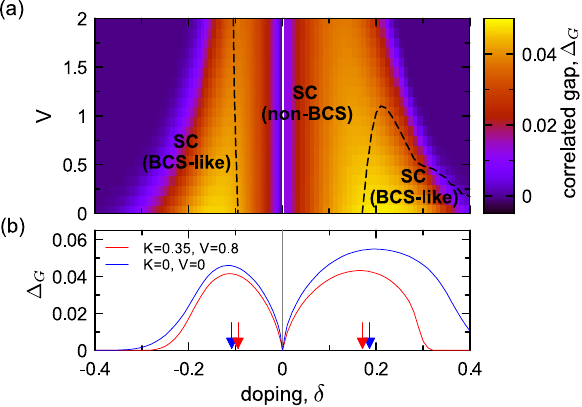}
\caption{Phase diagram for the full t--J--U--V model comprising both hole ($\delta < 0$) and electron ($\delta > 0 $) regimes of narrow-band filling. Both non-- and BCS--like regimes
are marked in each regime. The lower part illustrates the influence of intersite Coulomb (K) and correlated hopping (V) parts onto the superconducting dome shapes. The arrow marks correspond to the optimal doping points (after \cite{Spalek2022}).}
\label{Fig4}
\end{figure}

For detailed analysis of equilibrium properties of high-${\rm T_c}$ cuprates we refer to our detailed papers \cite{Zegrodnik2017,Dey2026,Spalek2022}, which comprises also determination of single-particle properties such as the Fermi-wave vector and chemical-potential dependence on the carrier concentration, etc. We turn now to a short determination of the dynamic excitations-paramagnons and plasmons (here we discuss in detail only paramagnons).

\subsubsection{Dynamic properties: Magnons and paramagnons}
To determine what is the relevant regime (BCS- or non-BCS-like) of hole concentration we have to compare change of the energy of single-carrier hopping ($z|t| n(1-n)$) with 
the maximal cost of breaking the singlet-pair energy ($J z n^2$, $z=4$ in the number of nearest neighbors). By equating those two energies we obtain the critical hole concentration $\delta = \delta_c \simeq 0.23$ for $J/|t| = 0.3$. 
For $\delta \lesssim \delta_c$ the carriers are strongly correlated, i.e., in the most of the superconducting dome (cf. Fig. 3 (b)). This simple argument illustrates again the general statement that the paired state is that of the strongly correlated quantum matter. Note that in this estimate
the atomic disorder created in real system by creating holes on $\ce{Cu^2+}$ sites is totally disregarded, so the Mott-localization effects, with an almost coinciding then antiferromagnetism appearance for $\delta \approx 0.05$, cannot be accounted for in such simple analysis.

With the above remark we ask now how to perform a more qualitative analysis of the additional fundamental aspects of the cuprates. The first of them is connected with an unusual feature of the single-electron structure coming from the angle resolved photoemission spectrum (ARPES) \cite{Fidrysiak2018,Hashimoto2014n}.
The typical result of the particle energy with respect to the Fermi energy
is presented in Fig. 5 (a), with characteristic kink in the dispersion relation. The kink is blurred in the measurements with a higher accuracy \cite{Hashimoto2014n}. From that we have drawn the conclusion that the two energy scales emerge from those measurements. Namely, the first of them concerns the particles with energies very close to the empty states are easier to excite than those lying deeper below, which are experiencing stronger correlations. In effect, the two linear portions $ \Delta \epsilon \cong \epsilon - \epsilon_{\rm F} = \hbar \nu_{\rm F} ( k - k_F) = \frac{\hbar k_{\rm F}}{m^*} \Delta k$ have different 
"Fermi velocities" $\nu_{F}$ (different effective masses $m^* $). 
Also, in Fig. 5 (d) the concentration dependence of the Fermi wave vector agrees well with the experiment. On the contrary, the doping dependence of quasiparticle weight factor $Z_{\rm nodal}$ in the $\Gamma-M$ direction (c) is reproduced qualitatively at best of the corresponding data trend. 
\begin{figure}[htb]
\centering
\includegraphics[width=0.9\textwidth]{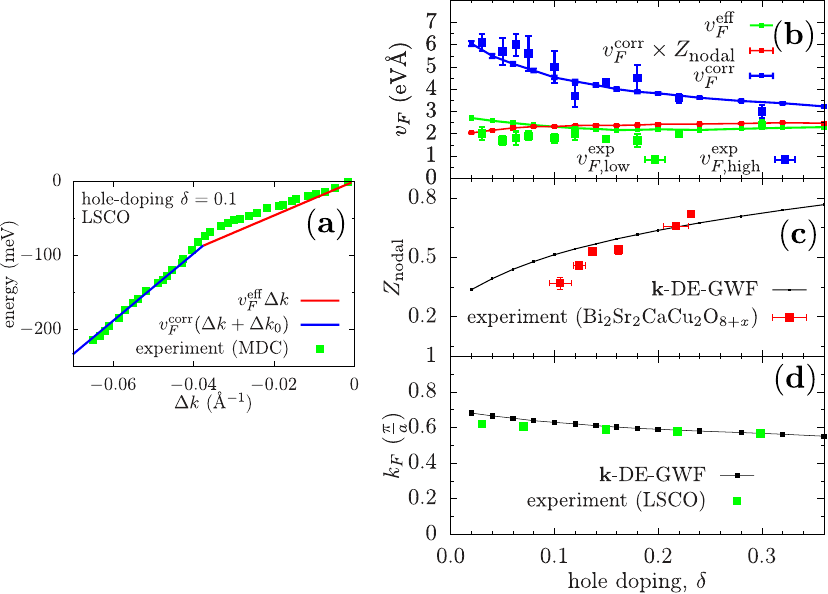}
\caption{(a) Schematic experimental ARPES results near the Fermi level in the nodal ($k_x = k_y$) direction for \ce{La_{1.8}Sr_{0.2}CuO4} as compared to the theoretical results (red line); (b)-(c) the resolved components of that dispersion relation and the spectral weights (c)-(d) for the two parts (for other details see Ref. \cite{Fidrysiak2018}.}
\label{Fig5}
\end{figure}
The second fundamental dynamic property of high-${\rm T_c}$ cuprates is the 
presence of collective magnetic excitations-paramagnons in the superconducting regime. Those dynamic fluctuations correspond to magnon excitations in magnetically ordered state, but  now have a nonzero lifetime even at $T=0$ due to absence of the long range order. On the other side, they represent a direct proof of an intrinsic role of the spin--spin (kinetic) exchange even though the ground state is now that of a superconducting spin-singlet. Starting 
from that presumption we have calculated the paramagnon spectrum in the Hubbard and t--J--U models within the time-dependent extended SGA approach, including leading $1/N$ contribution to the dynamic magnetic susceptibility \cite{Fidrysiak2020,Fidrysiak2021,Fidrysiak2021a}.
The exemplary results are collected in Fig. 6 (a)--(f) and compared with experiment.
Leaving the details to the original papers \cite{Fidrysiak2020,Fidrysiak2021,Fidrysiak2021a}  the main results can be summarized as follows: First, the upper (a,c,e) row represents the results obtained within our SGA + 1/N approach, whereas (b) and (d) represent those calculated in the standard random phase approximation (RPA). The SGA + 1/N results compare better with the data, compared to that obtained within random phase approximation (RPA) as one can expect, since the RPA approach is Hartree-Fock-type version with the time dependence included, but ignores completely the correlations.

\begin{figure}[htb]
\centering
\includegraphics[width=1.0\textwidth]{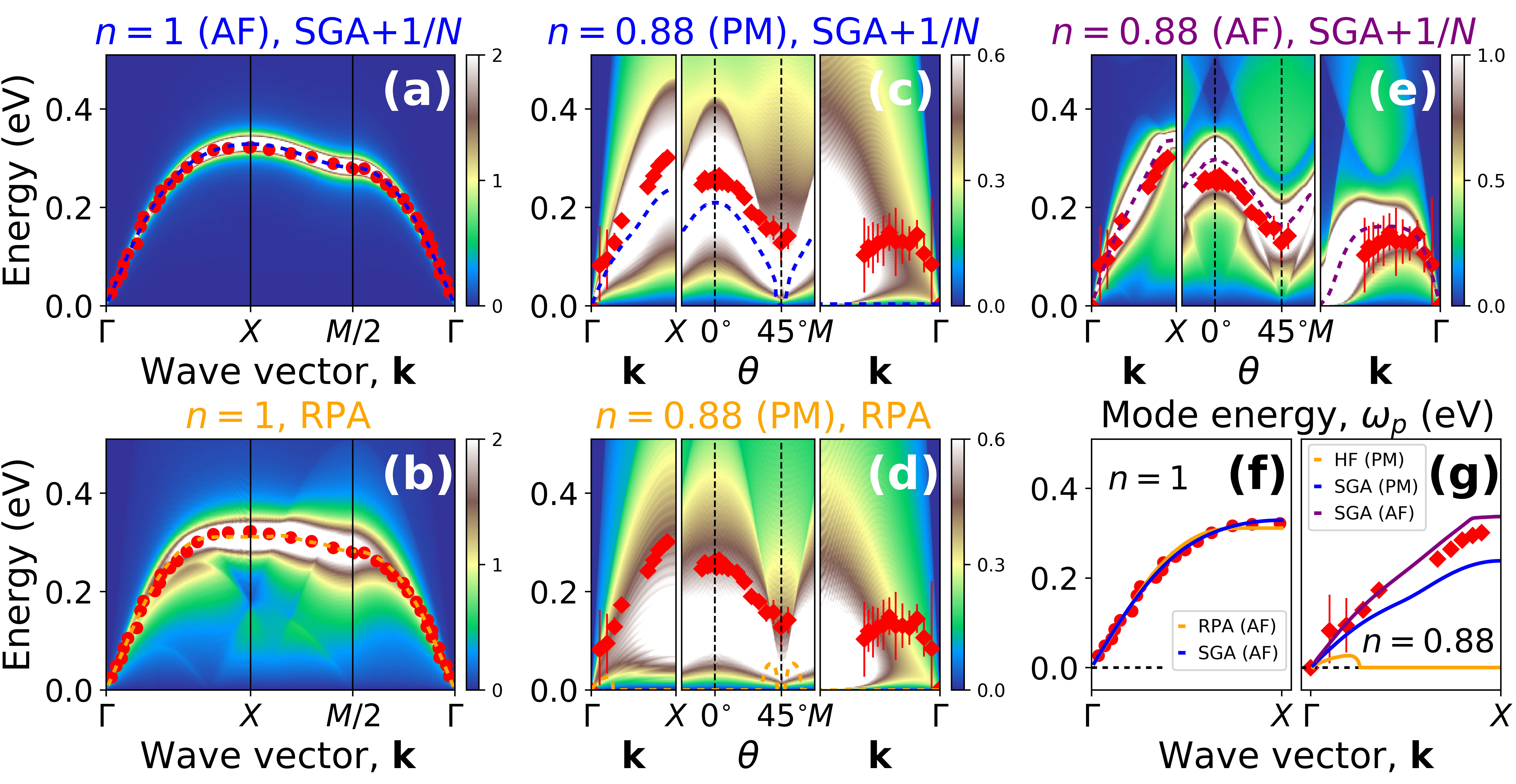}
\caption{(a)-(e) Spectra for the collective magnetic excitations comprising respectively: The magnon dispersion relation for the Mott insulator (a), paramagnon (c) and antiferromagnetic (AF) (e) spectra for strongly correlated metal for $\delta = 1 - n = 0.12$, calculated within SGA + 1/N approach; paramagnetic (b) and AF (d) spectra calculated within the RPA, analytic approximation for the spectra for $n = 1$ and $n = 0.88$ are provided in (f) and (g). 
For detail analysis see Refs. \cite{Fidrysiak2020,Fidrysiak2021,Fidrysiak2021a}}
\label{Fig6}
\end{figure}

From what has been shown in this section we can conclude that the t--J model (particularly with inclusion of intersite Coulomb interaction in some cases) describes the principal features of the cuprates as a two-dimensional strongly correlated electron system. The main unaccounted property in the whole DE-GWF approach in the quantitative manner is the appearance of pseudogap. Also, the influence of the atomic disorder on the persistence of the Mott insulating state at nonzero hole concentration ($\delta_c \sim 0.05$ for ${ \rm La}_{2-\rm x} {\rm Sr}_{\rm x} {\rm Cu O}_4$) and other high-$\rm{T_c}$ cuprates, remains to be clarified separately.

\section{ Heavy--fermion systems: Spin--dependent masses}
\subsection{ Basic characterization}

The second leading example of the strongly correlated quantum-matter systems are the systems with very heavy quasiparticles, i.e., with effective masses of $10^2 - 10^3$ times higher than  free electron mass $m_0$. In those systems the effective Fermi energy of the carriers can thus reach the value $\epsilon_F/k_B \sim  10-10^2 K$, i.e., unlike in ordinary metals we now have an access to study their properties in the achievable high-temperature regime. 
The canonical systems here are the cerium and uranium compounds such as $\ce{ Ce Al}_3, \ce{ CeCu_2 Si_2}, \ce{ UPt_3}, 
\ce{  U Be_{13}}$, \ce{Ce Co In_5}, and many others. In contrast to the Mott--Hubbard systems, here the simplest model description of those compounds is based on the
periodic Anderson model (PAM) composed of two subsystems of electrons: one consisting of initial atomic 4f electron/Ce atom in the case of Ce compound due to ${\rm Ce^{3+}}$ ions  located near energy $\epsilon_F$, of itinerant (uncorrelated) carriers. The typical model electronic structure is presented in Fig. 7. The most important feature is that the former electrons are strongly correlated and, at the same time, respectively, strongly hybridized with the uncorrelated carriers. Such a combination of the last two factors produces spectacular quantum fermionic liquid of heavy quasiparticles, particularly in the limit of almost integral valency of rare earth or actinide ions, e.g., for ${\rm Ce^{+3 - \delta}}$, with $\delta \sim 10^{-2}$, i.e., very close to the Mott localized state of 4f electrons (note that in the Mott localized state $\delta = 0$ strictly and the quasiparticles acquire infinite effective mass in the effective, renormalized by correlated hybridized narrow band).

\begin{figure}[htb]
\centering
\includegraphics[width=0.9\textwidth]{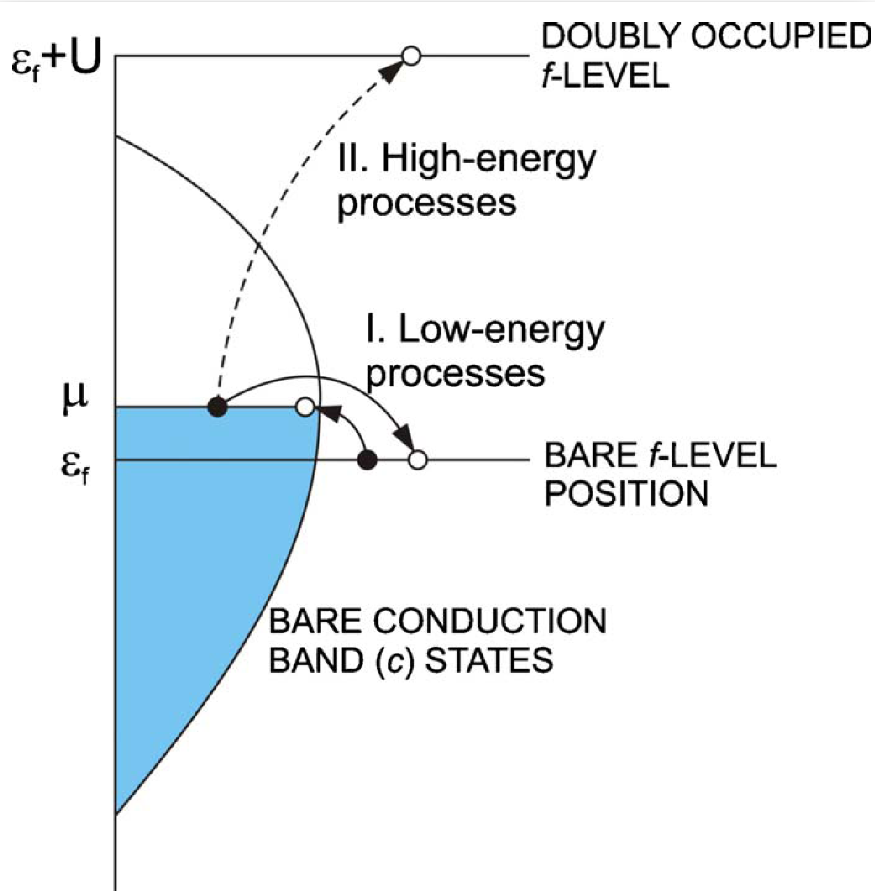}
\caption{Schematic representation of PAM, with bare atomic (f) and conduction (c) electrons hybridized with each other (marked by the arrows). In the limit of strong correlations the quantum mixing processes are divided into two parts, low (I)- energy processes and high (II)- energy processes. The processes (II) are virtual realized via virtual hopping $f \leftrightarrow c$ and lead to the Kondo-, superexchange-, and Działoszinskii-Moriya-exchange interactions of purely electronic origin \cite{Spalek1989,KadzielawaMajor2014}.}
\label{Fig7}
\end{figure}

Strictly speaking, in the Fig. 7 we characterize the situation in the strong-correlation limit. Explicitly, we have divided the hybridization 
(mixing) of the carriers with the strongly correlated (f-level) electrons into low- and high-energy processes. The former correspond to the hopping from singly occupied f level to the Fermi level and vice versa, whereas the high-energy express the hopping of carrier onto the localized level with formation of local doubly occupied f-state; those processes involve an additional energy $U \equiv U_{ff}$ of intraatomic (intrasite)
Coulomb repulsion between the f electrons. The principal point at this
stage is that U,  being the largest energy scale, leads to the virtual hopping processes only, as shown in Fig. 8 as exemplary second- and fourth-order contributions. The processes depicted in (a) lead to the Kondo interaction in the second (and higher) order, whereas those drawn schematically as (b) lead to the f-f superexchange, as well as, to the Dzialoshinskii--Moriya interaction of a purely electronic origin \cite{Spalek1989,KadzielawaMajor2014}. 
Explicitly, starting from PAM in the real-space representation
\begin{equation}
    H = \sum_{mn \sigma}\!\raisebox{1.4ex}{$\prime$} t_{m n} \hat{c}^\dagger_{m \sigma} \hat{c}_{n \sigma} + \epsilon_f \sum_{i \sigma} \hat{N}_{i \sigma} + U \sum_i \hat{N}_{i \uparrow} \hat{N}_{i \downarrow} + \sum_{i j} (V_{i m} \hat{a}_{i \sigma} \hat{c}_{m \sigma} + H.c.),
\end{equation}
and dividing the dynamic processes of type I and II (cf. Fig. 7) we obtain the following  effective Hamiltonian in the Fock space of lowest-energy states (i.e., with doubly f occupancies)

\begin{align}
    \hat{H}_{\rm eff} & \simeq \sum_{m \neq n, \sigma} (t_{m n} - \mu \delta_{m n}) \hat{c}^\dagger_{m \sigma} \hat{c}_{n \sigma} + \epsilon_f \sum_{i,\sigma} \hat{\nu}_{i \sigma} \nonumber \\
    & + \sum_{i , m ,\sigma} (V_{i m} (1 - \hat{N}_{i \overline{\sigma}}) 
    \hat{f}^\dagger_{i \sigma} \hat{c}_{m \sigma} + H.c) \nonumber \\
    & + \sum_{i , m} J_{i m}^{(K)} ({ \hat{\bf S}}_i \cdot \hat{\bf s}_m - \frac{\hat{n}_m \hat{\nu}_i}{4}) \nonumber \\
    & + \sum_{i \neq j, \sigma } J_{i j}^{(H)} (\hat{\bf S}_i \cdot \hat{\bf S}_j - \frac{\hat{\nu}_i \hat{\nu}_j}{4}) \nonumber \\
    & + 2 i \sum_{\langle m i \rangle \langle m j \rangle} J_{i j}^{(H)}
    ( 1 + \frac{n_f}{n_c}) \hat{\bf s}_m \cdot (\hat{\bf S}_j \times \hat{\bf S}_i).
\end{align}

\begin{figure}[htb]
\centering
\includegraphics[width=0.9\textwidth]{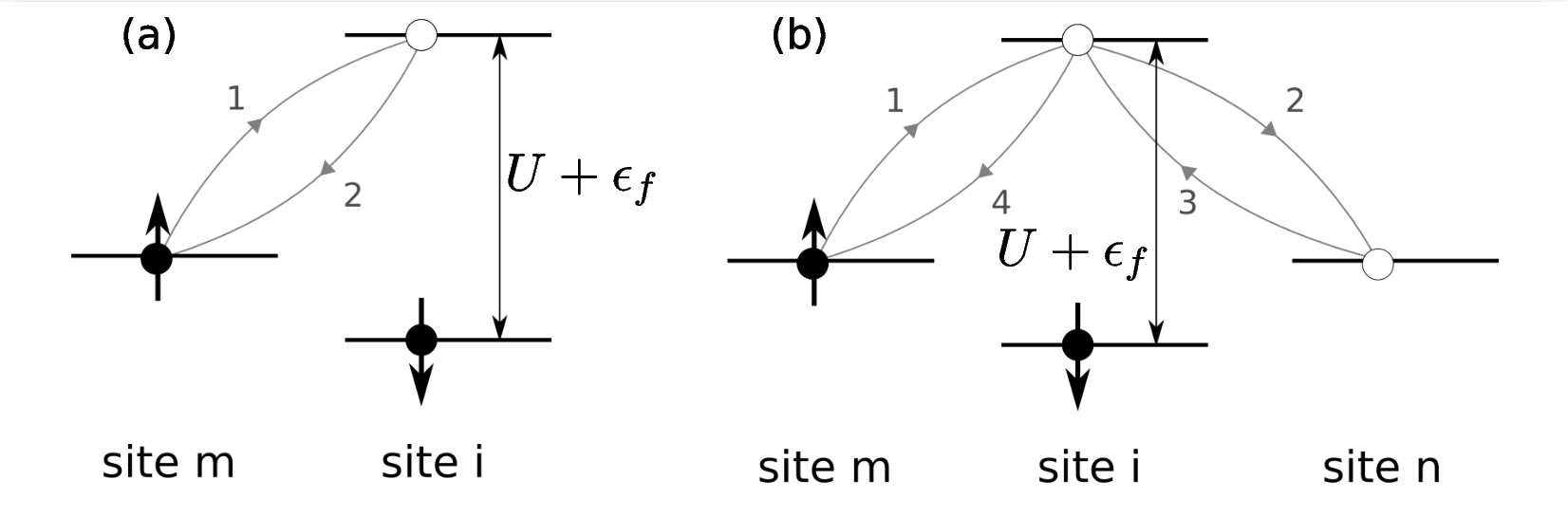}
\caption{Examples of  virtual hopping process contributing to the Kondo interaction in the second order (a) and the fourth-order term characterizing superexchange and Dzialoshinskii--Moriya exchange interactions. Note that to
obtain pure spin--spin interaction terms one has to have average over the 
degrees of freedom in the intermediate state.}
\label{Fig8}
\end{figure}

In the two above expressions $\hat{c}_{m \sigma } (\hat{c}^\dagger_{m \sigma })$ and
$f_{i \sigma} (f^\dagger_{i \sigma})$ represent annihilation (creation) operators of conduction and f electrons, respectively, whereas the first two terms represent single-particle energy of the former and latter respectively;
$\hat{N}_{i \sigma} \equiv \hat{f}^\dagger_{i \sigma} \hat{f}_{i \sigma}$, whereas the third and the last terms in (3.1) expresses the f-f intraatomic 
Coulomb interaction and the interorbital hybridization of individual-particle states. By contrast, the second and the third terms in (3.2) describe the projected atomic and hybridization energy parts with no double 
occupancies of the f-states, whereas the next three terms these represent 
respectively the Kondo, superexchange, and the Dzialoshinskii--Moriya interaction of purely electronic origin, as the $\{ \hat{\bf s}_m\}$ and 
$\{ \hat{\bf S}_i\}$ are the conduction- and free electron spin operators 
in the fermion representation (2.12). Finally, the corresponding effective exchange integrals can be estimated explicitly by averaging over intermediate-state fermionic parts and provide a ferromagnetic contribution to the Kondo exchange integral $J^K$, whereas the 
superexchange and Dzialoshinskii--Moriya counterparts are of the same order of amplitude \cite{Spalek1989,KadzielawaMajor2014}.

One very interesting feature should be noted. Namely, the magnitude of the
present novel type Dzialoshinskii--Moriya interaction can be of the same
magnitude as that of the antiferromagnetic kinetic exchange (superexchange) and may lead to a noncollinear magnetic ordering of the combined conduction-f electron type. That point requires a separate analysis elsewhere.

In summary, the effective exchange interactions here have been obtained within the same methodology as in the case of single band in Sec. 2. However, here we have now additional channels of virtual hopping of strongly correlated electrons via the conduction band(s). In effect, three 
kinetic-exchange interaction channels arise as exemplified by the Kondo,
superexchange, and Dzialoshinskii--Moriya parts. This circumstance, should lead
to a rich phase diagram. Moreover,
in the heavy-fermion limit ($n_f \to 1 - \delta,~ \delta \ll 1$), three 
physical regimes may be distinguished, depending on the ratio of hybridization magnitude V with respect to the atomic-level position $\epsilon_f$ of f-electrons, as illustrated schematically in Fig. 9, with overlapping properties across their boundaries. Each of these three physically distinct regimes have been discussed extensively in the literature.

\begin{figure}[htb]
\centering
\includegraphics[width=0.9\textwidth]{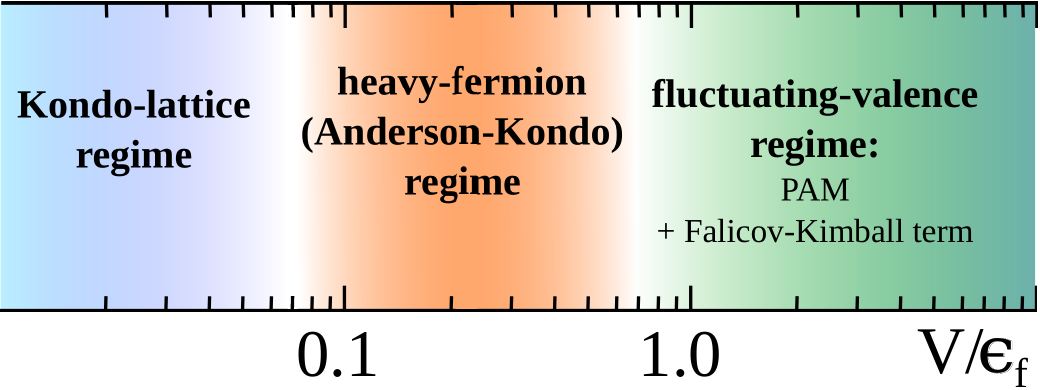}
\caption{Schematic representation of the various physical regimes evolving gradually and discussed in the literature. The x-axis is represented by the bare hybridization to f-level position ratio, $V/\epsilon_f$. Additionally, large-U limit is implicitly assumed.}
\label{Fig9}
\end{figure}

\subsection{Selected physical properties of heavy fermions}
The systems with very heavy electrons (i.e., with effective mass of carriers $m^* \sim 10^2 - 10^3 m_0$) are the systems with metallic properties on the edge of the Mott insulating state for f electrons
as a function of temperature. This is, as said already earlier,
because the Fermi energy of those carriers is in the range $T_F \equiv 
\epsilon_F/k_B \sim 10 \div 10^2 K$ and the  thermal disorder of the order
$T \sim 10 K$ is the source of the gradual transformation from low-T Fermi liquid to the metal of c-electrons localized f-electrons.

Throughout our work on heavy fermions we have taken the view that the f electrons in those systems are itinerant at low temperature even near the 
so-called Kondo-lattice regime, which should not be confused with the literal Kondo effect associated with the compensation of the individual magnetic moment of totally localized f electrons by that of uncorrelated conduction 
electrons. One should emphasize that such magnetic moment compensation is equally possible with having itinerant f-electrons \cite{Doradzinski1997,Doradzinski1998,SpalekDoradzinski1998}. In Fig. 10 (a) and (b) we show an exemplary situation with nominally one f electron (model
situation for nominally ${\rm Ce^{3+}}$ ion per site) and one conduction electron per site. In this figure we have analyzed the evolution of a partially compensated magnetic state as a function of magnitude of intraatomic hybridization to the bare bandwidth ratio $V/W \equiv V$ for the selected values of parameters
and the Lande factor $g=2$ and $g=6/7$ (that for ${\rm Ce^{3+}}$ ion). Different antiferromagnetic (AFM1, AFM2) and weak ferromagnetic phases have been analyzed as stable phases in the appropriate part of the phase diagram. 
\begin{figure}[htb]
\centering
\includegraphics[width=0.9\textwidth]{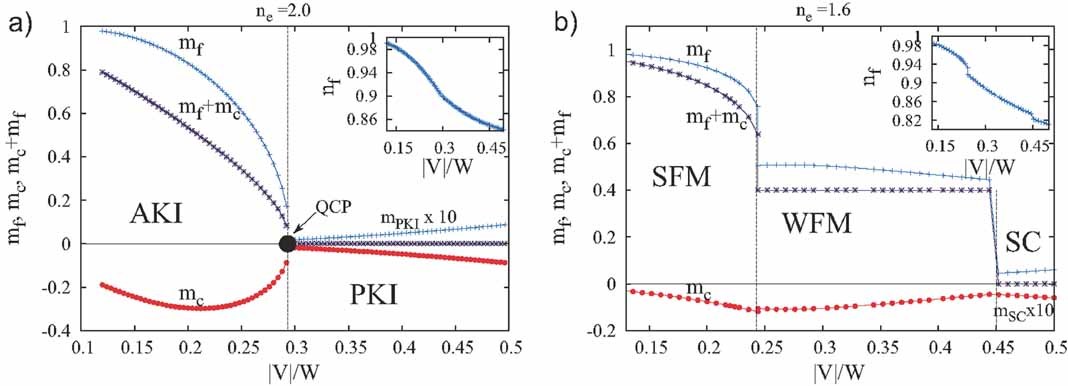}
\caption{ a) Phase diagram involving Kondo-insulator (KI) phases, both antiferromagnetic (AKI) and paramagnetic (PKI). Inset: f-level occupancy $n_f$. b) Same as a) but in the metallic state c with $n=1.6$ electron pair a-c of orbitals: weak (WFM) and strong (SFM) phases, as well as the superconducting (SC) unconventional phase. Red lines represent Kondo compensating moment ($m_c$), after \cite{Howczak2012}.}
\label{Fig10}
\end{figure}
The totally compensated magnetic-moment state is achieved at
higher hybridization magnitude of $|V| \sim 0.3$. To complete the picture we have
plotted in Fig. 10 b the situation for $n_e \equiv n_f +  n_c = 2.0$ electrons pair a-c pair of orbitals, where
a new type of insulator - the antiferromagnetic Kondo insulator
(AKI) has been detected for the first time which transforms into its paramagnetic 
correspondent (PKI) at the quantum critical point (QCP) appearing at $V \simeq 0.3$,
with totally compensated magnetic moments $m_f + m_c \equiv 0$ at and above 
QCP. In the inset, the f-level occupancy $n_f$ close to unity is shown at the same time. So the system then is indeed at the edge of f-moment localization ($n_f = 1$).

The above results show clearly that a totally magnetically compensated
state is possible even when the $n_c \approx n_f$, i.e., when their numbers
are comparable, unlike in the Kondo-impurity case when a single f-moment is screened by the Fermi sea of carriers. This magnetic moment compensation for the (Kondo) lattice is possible because here the f-electrons are itinerant. Hence, the f-moments 
are auto--compensating each other due to their fermionic nature. This happens in addition to the usual conduction-electron opposite 
magnetic moments of conduction electrons induced by the Kondo interaction. In result, both component moments disappear simultaneously and this effect is produced collectively by the Kondo-, fermionic-, and superexchange-effects in this almost localized Fermi liquid. All these features were obtained within the mean-field-like picture within the slave-boson
approach, which is equivalent to the later formulated and already discussed SGA method for high-${\rm T_c}$ cuprates and other single-band systems in normal phase.

\subsection{Complete phase diagram: Itineracy of f electrons and their pairing}
A more complete first diagram for heavy fermions requires still consideration of at least two features. First of them is the coexistence of almost integer occupancy 
of original f-states with almost compensated magnetic moments (see the preceding section) with very high density of states at Fermi level (heavy masses).  The last will be discussed in the next chapter. 
The second feature is the appearance of superconductivity of particles with those unusual heavy masses. This has been shown explicitly by proving that the quasi-jump of specific heat at the superconducting transition must involve them \cite{Steglich1979}.
This observation, as well as the observation of giant linear specific-heat 
coefficient \cite{Steglich1979,Andres1975} at low temperature, established the early enormous interest in those unique metallic-state materials.

In this overview, we limit ourselves to one particular topic, where we have made an original contribution. Namely, on the basis of our work presented
already in the preceding section, we have constructed fairly a complete phase diagram for those systems within mean-field picture of PAM on the plane $n-V$, comprising antiferromagnetic (AF), weakly-- (WFM); and strongly--ferromagnetic (SFM) phases, spin-singlet superconducting phase (SC), its coexistence regimes with AFM, as well as paramagnetic Kondo insulating (PKI) phase.  The results are summarized in Fig.11. 
The phase diagram illustrates a competitive nature for different phases of existing comparable energy scales (kinetic, residual hybridization, and exchange interaction). Those energies
are renormalized in the correlated states to become of that comparable magnitude in effect, leading in effect to such a complex mosaic of those phases.
\begin{figure}[htb]
\centering
\includegraphics[width=0.9\textwidth]{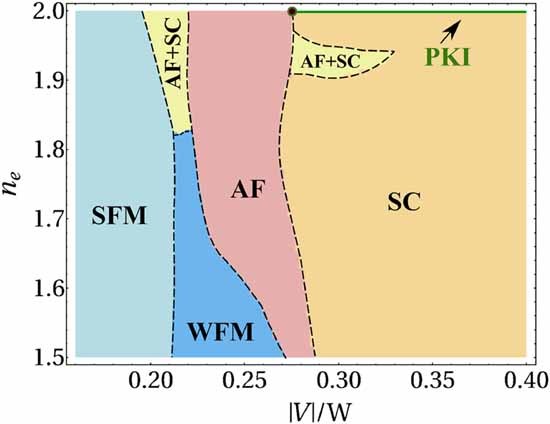}
\caption{ Overall magnetic phase diagram on the plane relative (intraatomic) hybridization magnitude $|V|/W$ - number of electrons per site ($n_e$). The symbols labeling the phases are explained in the text. The paramagnetic Kondo-insulator line (PKI) terminates the discontinuous-transition line with a quantum critical point. Superconductivity is of d-wave character. For details see \cite{Howczak2012,Howczak2012phd} and main text. The upper left-hand 
corner line below  QCP (solid point) may be that of coexisting antiferromagnetic Kondo-compensated superconducting phase.}
\label{Fig11}
\end{figure}

One another specific feature should be briefly mentioned here. Namely, the SC pairing considered here was the real space pairing, involving both that coming from the hybrid Kondo ($b^\dagger_{im}$) and superexchange ($B^\dagger_{ij}$) pairing parts defined now, respectively
\begin{align}
    \hat{b}^\dagger_{im} &\equiv \frac{1}{\sqrt{2}} (\hat{\tilde{f}}^\dagger_{i \uparrow} \hat{c}^\dagger_{m\downarrow} - \hat{\tilde{f}}^\dagger_{i\downarrow} \hat{c}^\dagger_{m\uparrow}) \equiv - (\hat{\bf{S}}_i \cdot \hat{\bf{s}}_m - \frac{1}{4}\nu_i n_m ),  \\
    B^\dagger_{ij} &\equiv \frac{1}{\sqrt{2}} (\hat{\tilde{f}}^\dagger_{i \uparrow} \hat{\tilde{f}}^\dagger_{j\downarrow} - \hat{\tilde{f}}^\dagger_{i\downarrow} \hat{\tilde{f}}^\dagger_{j\uparrow}) \equiv - (\hat{\bf{S}}_i \cdot \hat{\bf{S}}_j - \frac{1}{4}\nu_i \nu_j ),
\end{align}
where $\hat{f}_{i \sigma}$ projected Fermi operators are
\begin{equation}
    \hat{\tilde{f}}^\dagger_{i \sigma} \equiv  \hat{{f}}^\dagger_{i \sigma} (1 - \hat{N}_{i \overline{\sigma}}).
\end{equation}
In effect, the effective starting Hamiltonian, including the local pairing effects takes the form
\begin{align}
    \hat{H} = \sum_{m n \sigma} t_{mn} \hat{c}^\dagger_{m\sigma} \hat{c}_{n\sigma} + \epsilon_f \sum_{i \sigma} \hat{\nu}_{i \sigma} \nonumber\\
    + \sum_{i m \sigma} (V_{i m} \hat{\tilde{f}}^\dagger_{i \sigma} \hat{c}_{m \sigma} + H.c.) \\
    - \sum_{i m n} J^K \hat{\tilde{b}}_{im} \hat{\tilde{b}}_{i n} - \sum_{i j} J^H_{i  j} \hat{\tilde{B}}^\dagger_{ij} \tilde{B}_{ij}.
\end{align}
Note that the last two terms provide  lowering of the energy due to the local pairing. Their pairing operator representation contains thus, in an obvious manner, to the contribution for the antiferromagnetic state, so that this is why coexistence appears. Such a formal formulation points to the common (magnetic) origin of high-${\rm T_c}$ and heavy-fermion superconductivity \cite{Spalek1988a}. The proof of this universality is yet to be explored further and consequently, tested experimentally. 

In summary, our analysis of real-space spin-singlet superconductivity, with two gaps, due to the Kondo- and superexchange-pairing, respectively \cite{Zegrodnik2019}, relies on a direct analogy between the pairing in high-${\rm T_c}$ cuprates (in single-
band- and three-band \cite{Zegrodnik2019} cases) and this is one of the crucial features that distinguishes the strongly correlated quantum liquids from other metallic systems. The difference between them is that the high-$\rm{T_c}$ systems evidently represent examples of non-Landau quantum fermionic liquid, whereas some of the heavy-fermion liquids can be regarded as (almost localized) Landau Fermi liquid systems, albeit with nonstandard, quantum critical behavior in their low-temperature regime. However, this last fundamental phenomena will not be discussed in detail here. Instead, we briefly elaborate on the spin-dependent heavy masses which, in our view, possess also a potentially equally fundamental importance, as we discuss next.

\subsection{ Spin dependent masses: (In)distinguishability of strongly correlated heavy electrons}

One of the most interesting concepts invoked in my group was the introduction
of spin-dependent effective masses \cite{Spalek1990a,Korbel1995,Spalek2006,Spalek2006a}. This concept is specific for strongly correlated fermions and is due to the spin-dependent renormalization of the hopping part in any model with dominant Hubbard (intraatomic Coulomb) interaction. The simplest treatment relies on noting that when interaction $U/W \gg 1$ and the local double occupancies $\langle n_{i \uparrow} n_{i \downarrow} \rangle \equiv d^2 = 0$. Then, the hopping probability reduces to $\langle 0 | \hat{a}^\dagger_{i\sigma} \hat{a}_{j \sigma}|0 \rangle_{U\to\infty} \cong (1 - n)n_{\sigma}$. This probability in the Fermi-liquid limit can 
be estimated as $\langle\hat{a}^\dagger_{i \sigma} \hat{a}_{j \sigma}\rangle_{FL} \simeq n_\sigma (1 - n_\sigma)$.
In effect, regarding the hopping in $U\to\infty$ as a renormalized Fermi-liquid correspondent we obtain that
\begin{equation}
    \langle\hat{a}^\dagger_{i \sigma} \hat{a}_{j \sigma}\rangle_{U\to\infty} = q_\sigma \langle\hat{a}^\dagger_{i \sigma} \hat{a}_{j \sigma}\rangle_{FL}
\end{equation}
We see immediately that if we define the bare mass in the tight binding approximation as 
\begin{equation}
    m^* = \frac{\hbar^2}{|t| a^2},
\end{equation}
where $|t|$ is nearest neighbor hopping and $a$ is the lattice constant,
then in the strongly correlated state
\begin{equation}
    m^* \to m^*_{\sigma} = \frac{1 - n_\sigma}{1 - n} m^* \equiv q^{-1}_\sigma m^*.
\end{equation}
This, in turn, means that:
\begin{enumerate}[label=(\roman*)]
\item the effective mass enhancement is explicitly spin--direction dependent in the magnetically polarized medium, and
\item the spin-minority $(\sigma = -1)$ mass is divergent in the Mott-insulator limit $n=1$, when the spin-majority ($\sigma = \pm 1$) mass is reduced to the bare band mass $m^*$ as $n_{\uparrow} = n$.
\end{enumerate}
 Both (i) and (ii)  were subsequently observed \cite{Sheikin2003,McCollam2005,Shishido2018}. The case (ii) is difficult to observe in full,, since usually the first-order transition takes place before the actual singular point $m = n \delta_{\sigma_\uparrow \sigma_\uparrow}$ is reached, see, e.g., the case of liquid ${\rm ^{3}He}$ crystallization \cite{Wysokinski2014} and in general \cite{Semeniuk2023}.

 In Fig. 12 we plot schematically the band splitting  evaluation in the magnetically polarized state.
 It should be noted that, in addition to the ordinary Zeeman spin-subband splitting, in nonzero applied magnetic field, we have also predominant spin-asymmetric band narrowing due to the spin-dependent band narrowing factor $q_\sigma$. The second effect leads also to the metamagnetic behavior of the magnetization curve and is caused by a highly nonlinear behavior of $q_\sigma$ on spin polarization $m\equiv \langle n_{i\uparrow} - n_{i \downarrow} \rangle$. This behavior changes very rapidly when the band filling is away from the situation very close to the half-band filling. This particular behavior is illustrated in Fig. 13 ab, where respectively the magnetization curve (a) and spin-split mass have been drawn. Note that in the saturation state ($m \to n$) and $m^*_\uparrow = m^*$, i.e., the bare band mass is achieved for the majority electrons and this may be tested experimentally. 
 Such test would allow for a separate determination of the bare (band) $m^*$ and the correlated (spin-dependent) mass $m_\sigma$.
\begin{figure}[htb]
\centering
\includegraphics[width=0.9\textwidth]{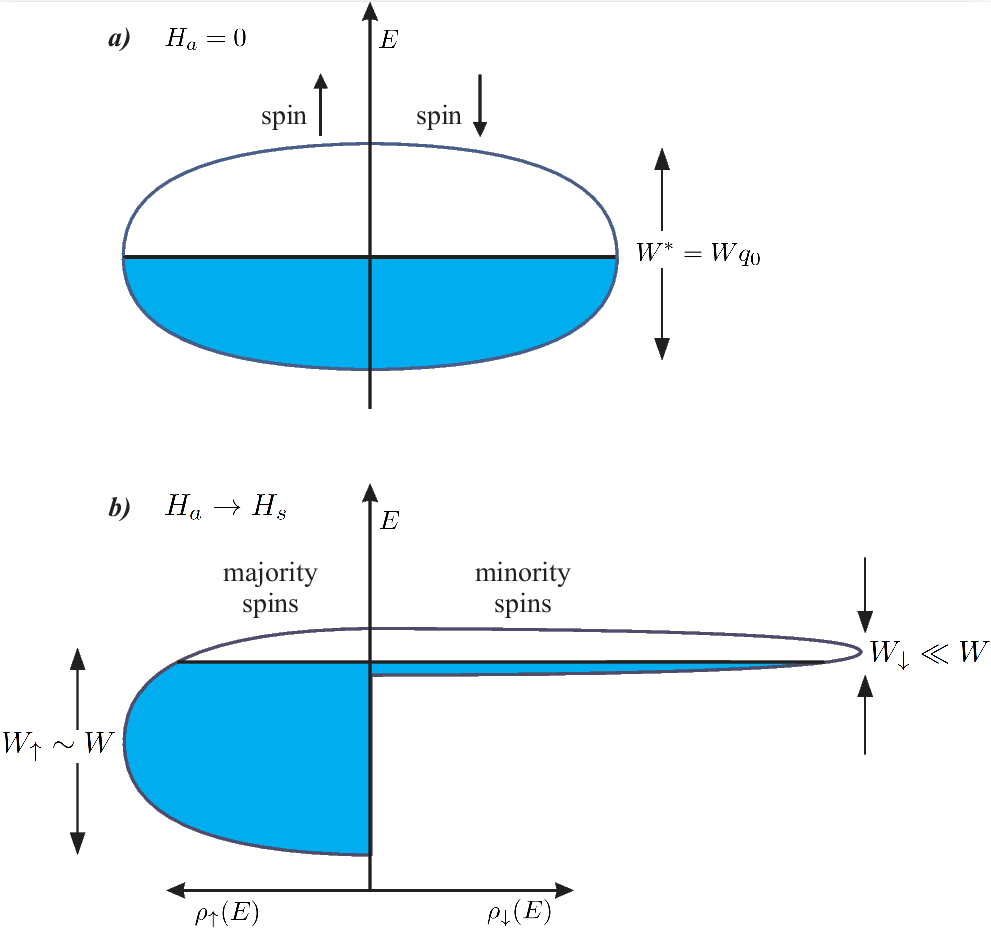}
\caption{Schematic illustration of the spin-subband picture of narrow band in the absence (a) of the applied field ($H_a = 0$) and that of spin-asymmetrically distorted (b) when $H_a \neq 0$. The latter effect is highly nonlinear in increasing $H_a$ and leads to the strong metamagnetic behavior of almost localized electrons. The Zeeman-field spin splitting is usually negligible near the metamagnetic transition point.}
\label{Fig12}
\end{figure}

\begin{figure}[htb]
\centering
\includegraphics[width=\textwidth]{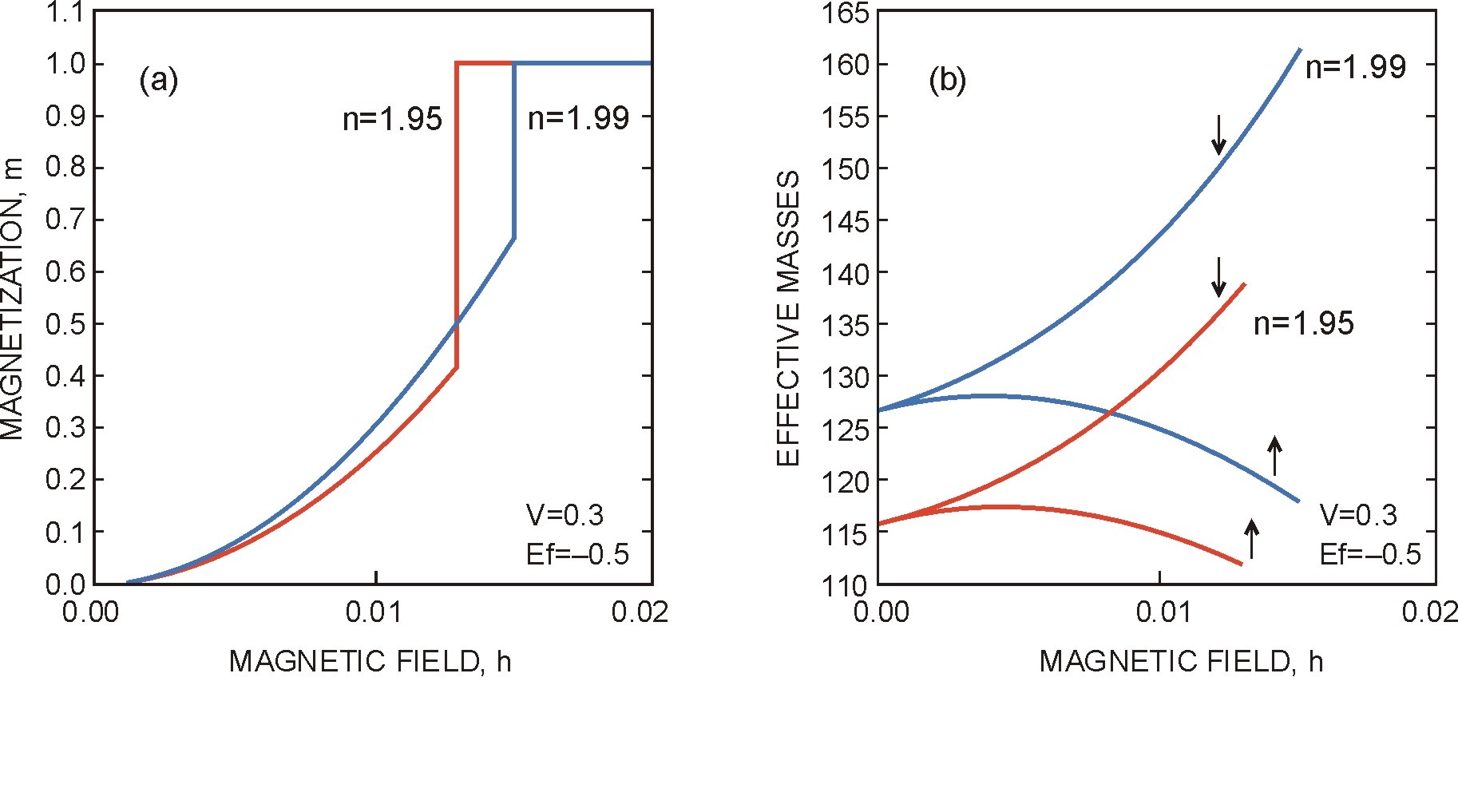}
\caption{The metamagnetic behavior (a) and the spin-split structure in the Anderson-lattice-model situation. The total number of electrons per site is specified in each case, as well as the relative magnitudes of intraatomic hybridization and the position $\epsilon_f$ of the f-level with respect to the zero field Fermi-level position (after \cite{Citro1999}).}
\label{Fig13}
\end{figure}

 The spin-dependent masses have been observed experimentally in the form of spin-resolved de Haas--van Alphen oscillations \cite{Sheikin2003,McCollam2005,Shishido2018}. If the mass difference is not too large then one should observe only the quantum beats of the two components with slightly different frequencies due to the corresponding cyclotron-resonances difference
 \begin{equation}
     \omega \to \omega_\sigma = \frac{e}{2 m^*_\sigma c} \simeq \frac{e}{2 m_0 c} \cdot \frac{1-n}{ 1 - n_\sigma} \equiv \omega_0 \frac{1 - n}{1 - n_\sigma},
 \end{equation}
 where $n_\sigma \equiv \frac{n}{2} + \sigma \frac{m}{2}$ and $\omega_0$
 is the frequency in the magnetically saturated state. 

 One important feature of the observability of the spin-split masses should be noted. The effect should be present in both single-band and hybridized correlated-band systems. So far, it has been observed only in the latter (heavy-fermion) systems. Then, since the bands there are indeed very narrow ($q W \lesssim 10-10^2 K$), the influence of the much weaker applied-field energy ($\mu_B H_a \sim 10-30 T \sim 1 K$) becomes appreciable only in that limiting situation. But even then, the effect is relatively 
 small \cite{Shishido2018}. The narrower effective heavy-quasiparticle (f-band)
 effect  is more pronounced. To illustrate this fact we would like to quote \cite{Kurleto2021} a direct determination of the zero-field heavy mass via ARPES for \ce{CeCoIn_5}, for which the spin-dependent masses have been observed \cite{McCollam2005}. The results are displayed in Fig. 14. As we can see, the system is of multiband character and the masses (in zero field) vary in the interval $m^* = 30-130 m_0$.
\begin{figure}[t]
\centering
\includegraphics[width=\textwidth]{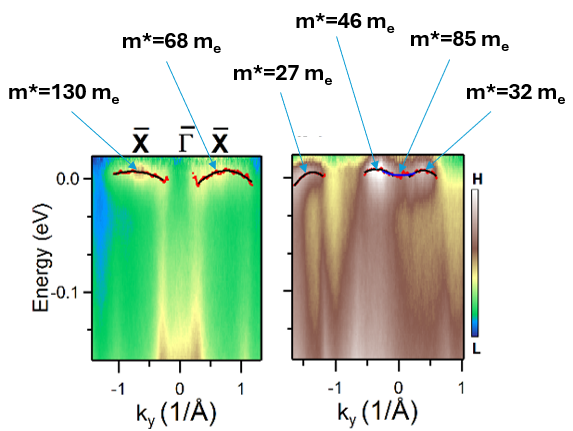}
\caption{Exemplary ARPES spectrum \cite{Kurleto2021} for \ce{CeCoIn_5} near the Fermi level fitted with a parabolic dispersion relations and corresponding effective masses marked (courtesy of P. Starowicz). For details see Ref. \cite{Kurleto2021}.}
\label{Fig14_1}
\end{figure}
 We turn next to the directly fundamental quantum- and statistical-mechanical meaning of some of the strong-correlation effects in the general terms.
 \section{Selected general features of other correlated systems}
\subsection{(In)distinguishability of spin-masses (quasi)particles in an almost localized Fermi liquid}
As said above, there is, on one hand, possible to extract the bare (band) mass $m^*$ in correlated systems from their (spin-split) renormalized correspondents $m_\sigma$ in the fully (partially) polarized states. However, in the latter state with spin partial magnetic polarization below the metamagnetic jump to the saturated states, cf. Fig. 13 a), the particles with moment $\sigma = \uparrow$ and $\sigma = \downarrow$ can be regarded {\it as distinguishable in the quantum-mechanical sense}. This is because they have remarkably different effective masses, what makes those two Fermi-liquid subsystems separable. This observation of fundamental importance is being analyzed by us in detail  at the moment \cite{Spalek2026n}.
The effect associated with the distinguishability via different masses may not always be small, as the mass difference $m^*_{\downarrow} - m^*_{\uparrow}$ may easily reach or even exceed the average-mass value, $\frac{1}{2} (m^*_\uparrow + m^*_{\downarrow})$. The situation in the present case is as follows. We start from the paramagnetic phase and the heavy quasiparticles are indistinguishable, but may not obey F-D statistics even when the at applied field $H_a = 0$.
Then, in the low-field regime, where magnetic moment $m\equiv n_\uparrow - n_\downarrow$ increases continuously (cf. Fig. 13 (a)), the particles become
{\it distinguishable} in a clearer form as ($m^*_\downarrow - m^*_\uparrow$)
increases and then, when magnetization jumps to the saturation state $m\simeq n$, they become again indistinguishable and obey Fermi--Dirac statistics, since the interaction $U n_\uparrow n_\downarrow$ is then switched off. In principle, this type of analysis may have a basic importance
for testing of the interplay between behavior of strongly correlated 
systems in relation to their fundamental quantum-mechanical statistical properties. The details
of this research is deferred to a separate paper \cite{Spalek2026n}.

\subsection{Advanced methodology of precise treatment of correlations: Exact diagonalization { \it ab  initio}  approach (EDABI)}
The discussion of the high-temperature superconductivity in section 2 was based on the diagrammatic expansion for the variational wave function (DE-GWF),
which in turn started from the statistically consistent mean-field approach (SGA) \cite{Spalek2022}.
All those considerations start, from the microscopic model (Hubbard, t--J, 3-band, PAM, etc.) with small number of
parameters, which in turn are defined in terms of single-particle Wannier functions that appear in them in an implicit form. The methods which combine both the determination of single-particle wave functions (${\rm 1^{st}}$ quantization of the problem) and the interparticle correlations (${\rm 2^{nd}}$ quantization aspect of the problem) are more difficult to tackle and usually carry out with their implementation deficiencies such as double counting of interaction (cf. DFT+U, DFT+DMFT) \cite{Pavarini2022n}, though 
in some formulations such deficiencies are to be removed.

In our first formulation of \textbf{E}xact \textbf{D}iagonalization \textbf{Ab} \textbf{I}nitio(EDABI) method we have referred to the original formulation of many-particle Hamiltonian in the Fock space \cite{Fock1957book}, where it is rigorously decomposed in one- and two-particle parts, $\hat{\mathcal{H}}_1$ and ${\hat{\mathcal{H}}_2}$, respectively. In such formulation the single-particle (Wannier, molecular) basis is defined with the help of $\hat{\mathcal{H}}_1$ part only.
However, as we are interested in (strongly) correlated systems, those functions should be readjusted (renormalized) in the resultant correlated state, when $\hat{\mathcal{H}}_2$ is included in analysis. In our model considerations, limited so far to single-band Hubbard model or the simplest molecular systems, we start from Slater atomic-function basis of variable size (Bohr-orbit-type with adjustable Bohr orbit size), which is explicitly determined by the minimization of the total energy in the final correlated state. In this 
manner, the single- and multiple-particle aspects are treated on the same footing, but without any double-counting of the interaction. Moreover, the 
energy of the resultant correlated state may be sometimes obtained from the exact diagonalization (Lanczos method, Lieb--Wu solution for one-dimensional Hubbard method, etc.). This last feature of the analysis limits severely the applicability of our method to realistic systems. Nonetheless, it is valuable to have a handful of rigorous and truly first-principle results for the testing of more complicated methods of approach.

\begin{figure}[h]
\centering
\includegraphics[width=0.9\textwidth]{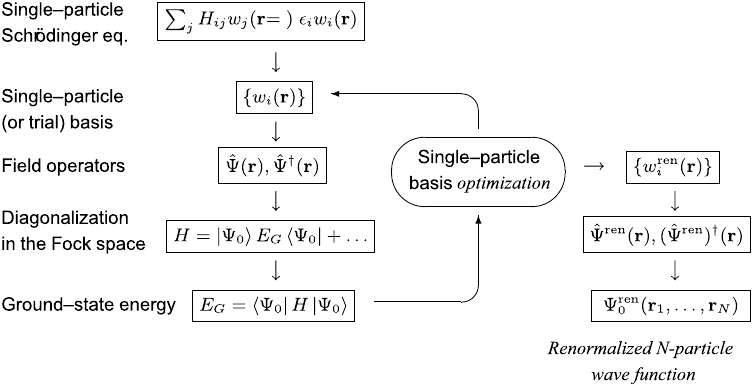}
\caption{Flowchart describing the scheme of the EDABI method. For details see main text. When selecting \textbf{ad hoc} the single-particle set, the topmost block should be disregarded. The renormalized many-particle wave function $\Psi_{0}^{ren}(\textbf{r}_1,\ldots, \textbf{r}_N) $ is explicitly constructed for N-particle systems by transformation from its second-quantization representation, as discussed elsewhere \cite{Hendzel2026}.}
\label{Fig14}
\end{figure}

In Fig. 15 we provide a flowchart of exemplary computational scheme within EDABI method. Exemplary basic results are discussed to some detail in the next subsection.

\subsection{Exact results for correlated nanochains}
As first application of our method we discuss the physics of nanochains composed of H atoms, up to 18 atoms in linear-chain or ring configurations \cite{Spalek2020}. One of the most important results is the evolution of exact statistical distribution function $n_{\bf{k}\sigma}$ as a function of interatomic 
distance $a \equiv |{\bf R}_{ij}|$. The result is displayed in Fig. 16
for small interatomic distance $a$, equal to the size of the single Bohr atom ($2 a_0$) we observe an almost typical slightly modified Fermi--Dirac distribution with lower quasimomentum $k \gtrsim k_F$ nonzero occupancy due to the repulsive Coulomb 
interaction (Fermi-liquid type correction). In contrast, for the relative large interatomic separation, ($a/a_0 \gtrsim 5$) the distribution is continuous
and centered about half occupancy.
\begin{figure}[h]
\centering
\includegraphics[width=0.9\textwidth]{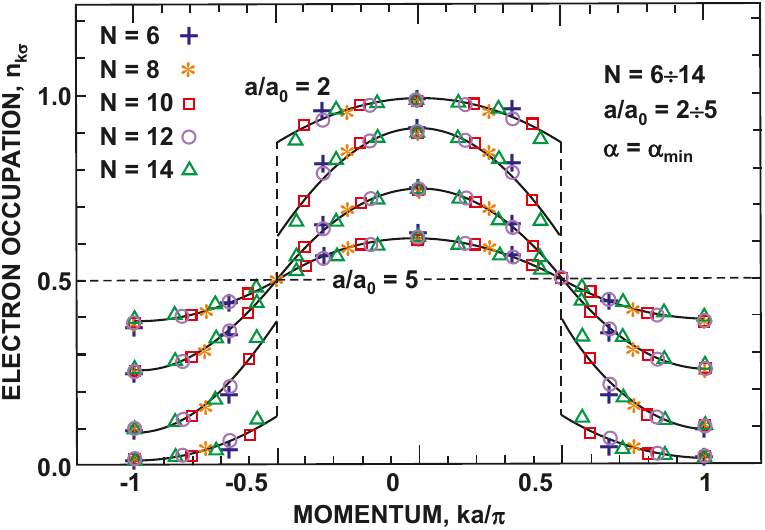}
\caption{Statistical distribution function $n_{\bf{k}\sigma}$ for $N=6-14$ hydrogen 
atoms obtained from exact diagonalization ab initio (EDABI) method. The evolution from
Fermi--Dirac like to continuous case with increasing interatomic distance $a/a_0$ signals a gradual transformation of itinerant to localized atomic states of the electrons.}
\label{Fig15}
\end{figure}
Such a situation with a smooth and almost
constant $n_{k \sigma} \approx \frac{1}{2}$ distribution we regard as 
a clear sign of localization of carriers on their parent atoms. 
In other words, it signals the onset of the {\it Mott localization} for the 1s-type electrons on the parent atoms. Note that this type of localization 
cannot be regarded as a Wigner crystallization, as in the former situation we start with atomic states, out of which an interacting fermionic liquid
evolves with decreasing $a/a_0$, not by the freezing out of the electron gas
having a neutralizing uniform background. In other words, the two -- Mott and Wigner localizations--, are complementary phenomena, with the Mott (or Mott-Hubbard) picture being physically proper representation of correlated liquid behavior on a lattice.

\begin{figure}[htb]
\centering
\includegraphics[width=0.9\textwidth]{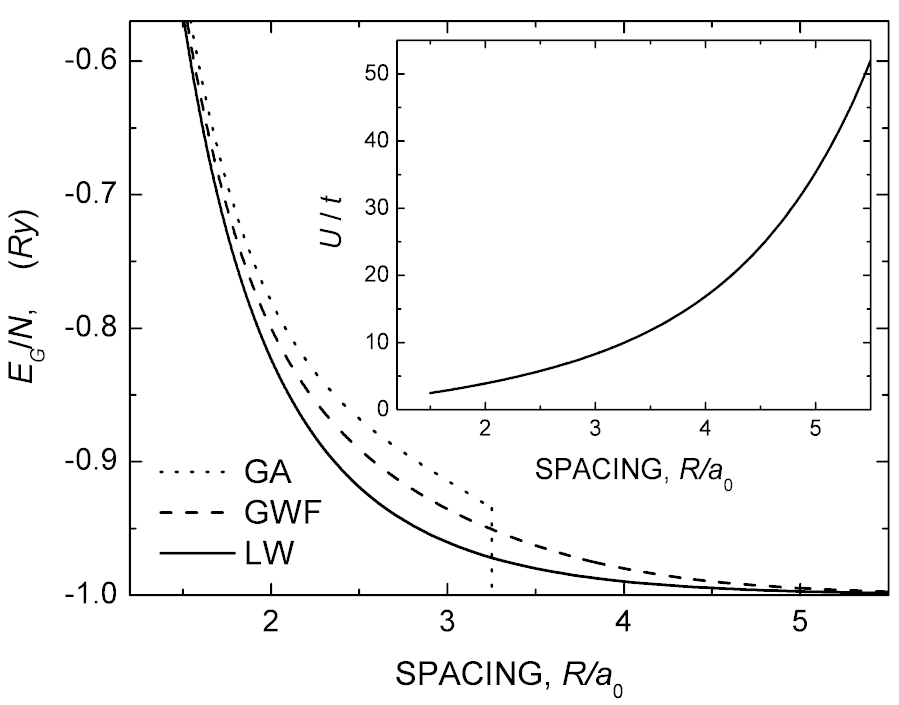}
\caption{Variational of the ground state energy (per site) versus $R/a_0$ within the three schemes exact Lieb--Wu (LW), full Gutzwiller wave-function (GWF), and the mean-field (SGA) solutions. Inset: Explicit $U/|t|$ ratio vs. $R/a_0$; this feature can be obtained from the explicit application of EDABI. For details see \cite{Kurzyk2007,Kurzyk2008,Spalek2010}}.
\label{Fig16}
\end{figure}

As a second example of our method application we have discussed the exact 
Lieb--Wu solution \cite{Lieb1968,Lieb2003} of the Hubbard model by supplementing it with the single-particle wave-function optimization \cite{Kurzyk2007,Kurzyk2008}. In this way, we can discuss the solution as a function of the relative lattice spacing $R/a_0$. 
In Fig. 17 we have plotted the double occupancy probability $d \equiv \langle n_{i \uparrow} n_{i \downarrow} \rangle$ as a function of interatomic distance $R/a_0$ and have compared three solutions: Exact Lieb--Wu (LW),  full Gutzwiller wave-function (GWF), and the simplest Gutzwiller (GA) - results \cite{Kurzyk2007}. 
One can see that the first two solutions provide a smooth evolution with
the increasing spacing, whereas (mean-field) SGA solution provides a critical lattice spacing, at which $d^2 \equiv 0$ and signals the continuous
transition to the Mott insulating state for $R/a_0 > 3.25$.
Parenthetically, the exact solution of Lieb and Wu yields only Mott insulating solution for all $R/a_0$ at $T=0$. Here we have included \textbf{all} Coulomb interactions, so the results do differ. Also, the wave functions $\{w_i{\bf{\sigma}}\}$ have been adjusted in the correlated state; i.e. $|t|$ and U differ with changing $R/a_0$. Inset illustrates the behavior of $U/|t|$ ratio vs. $R/a_0$. Otherwise, in the standard approach $U/|t|$ would be the microscopic parameter as  available if there was no simultaneous 
computation of the single particle wave function.

Later, the approximate analysis within the EDABI method have been carried out for an extended  Hubbard model with the charge neutrality (i.e., presence of positive ${\rm H^+}$ lattice) assured. A critical behavior of the single-particle wave function was observed near the Mott transition 
for all three cubic lattices (sc, bcc, and fcc) \cite{Spalek2010}.
As far as we are aware of, no such effect has been analyzed, yet alone observed experimentally. 

We now turn to our most recent analysis of the exact analytic solution of Heitler-London solution of \ce{H_2} molecule.

\subsection{Digression: Strong correlations in simple molecules and atomicity in the chemical bonding: ${\rm H_2}$ molecule example}

As the last and elucidating example of applicability of the concepts of strong correlations presented so far we consider the simplest molecular system - ${\rm H_2}$ molecule. Our part of the involvement in this
topic \cite{Hendzel2022} originated from the author's exact solution of the Heitler-London model of ${\rm H_2}$ \cite{Spalek2007a}. In the original Heitler-London version
of the model, only 1s-hydrogen orbitals have been taken into account without their size renormalization in the molecular state and subsequently, 
the ground state energy calculation and the whole solution analysis carried
out in the Hartree--Fock approximation. It turns out that no such simplifications are necessary to make and a rigorous analytic solution 
can be obtained in an analytic  form by combining ${\rm 1^{st}}$ and ${\rm 2^{nd}}$ quantization aspects of the problem. We concentrate here only on one novel aspect of the
solution, namely on the concept of {\it atomicity}. This particular topic appeared once we have noticed that the standard discussion of the covalency leads to an unphysical property of the covalency with the increasing interatomic distance. Namely, the covalency is maximal at infinite interatomic distance, a clearly wrong result.

To illustrate our results obtained for our exact solution of the extended Heitler-London model, we have plotted in Fig. 18 a-d the electron density profiles with the increasing interatomic distance $R$ from $R = 1 a_0~(a)$ 
to $R = 4 a_0 ~(d)$. We see the gradual separation into two well defined maxima centered at the proton locations. By inspecting these evolution 
we have shown that a well defined Mott and Hubbard criteria of localization are fulfilled simultaneously $R = R\simeq 2.3 a_0$, well above the equilibrium 
bond length $R \simeq 1.43 a_0$ bond.
\begin{figure}[t]
\centering
\includegraphics[width=0.9\textwidth]{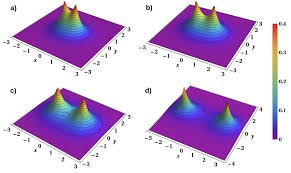}
\caption{The electron density profiles for \ce{H_2} molecule at different interatomic
distance $R/a_0$: $R=a_0$ (a), $1.43 a_0$ (b), $2.3 a_0$ (c), and $4 a_0$ (d), obtained within the exact solution of the extended Heitler-London model \cite{Hendzel2022}. A gradual separation of atoms is clearly visible and allows for an analysis degree of \textbf{atomicity} in the bonding state for details see \cite{Hendzel2022, Broclawik2023}. }
\label{Fig17}
\end{figure}
This means that for $R \ge R_{\rm Mott }$ the Coulomb repulsion between
electrons exceeds the kinetic energy \cite{Hendzel2022,Broclawik2023}. In this large-separation
limit the correlated two-particle states gradually transforms into two separate
atomic 1s states. As already has been said, the standard approach provides an unphysical features that the covalency increases further for $R>R_{\rm Mott}$ and reaches its maximum of unity in the atomic limit $R \to \infty$. To correct that we have introduced the concept of atomicity, expressed at the state when formation of single-particle molecular states is disregarded in the whole
analysis.

In effect, the covalency is reduced; instead, the true covalency, ionicity, and atomicity,
 can be properly defined \cite{Hendzel2022,Broclawik2023}, all as a function of $R/a_0$.  The proper behavior in the asymptotic limit $R > R_{\rm Mott}$ of all the bonding 
 factors, together with that of atomicity have a proper behavior in the asymptotic limit $R\to\infty$. This is the first such result of the kind in the literature  \cite{Broclawik2023}. Connection of this evolution with the entanglement correlation of the electrons will be discussed elsewhere \cite{Hendzel2026}.

\section{Concluding remarks}
In this overview I have touched upon some aspects of my research on strongly correlated electron systems, starting from the kinetic exchange interaction
Hamiltonian (1976) for the itinerant electrons (t--J model and its extensions) as a generalization of the effective spin Hamiltonian for the
Mott insulators. These ideas have led to an exact expression of 
the pairing operators in real space (1988) and their subsequent applications to the high-temperature superconducting cuprates. The research in high-${\rm T_c}$ cuprates was summarized in review article (2022). However, that required a statistical correction to the original resonating-valence-bond or Gutzwiller approach in the form of SGA (2010) and, eventually, to the systematic DE-GWF (2017) approach, within which some principal properties of the cuprates can be described in a semiquantitative manner. As a helpful practical tool was also the proposal to use the t--J--U version of the model in practical calculations, as it does not involve a strict elimination of local double--site occupancies, but is accounted for automatically when $U \to \infty$.

In this paper I have omitted details of my work creating the first thermodynamical 
mean-field theory of the Mott--Hubbard transition (1986--1989). Nonetheless, the associated with this research on properties of an almost localized Fermi liquid, the work on the concept of spin-dependent effective masses (1990--2007) has been included. 

Finally, I have mentioned the work on strongly correlated nanophysical chain and simple molecules (2000--2003).

I have also omitted the recent works on quantum fluctuations (paramagnons)
in the cuprates, as it is reviewed separately \cite{Fidrysiak2026}. At present, our
group is also working on nonstandard statistics; this is briefly elaborated also separately in \cite{Spalek2026etal}, as well as entanglement of correlated electrons in the hydrogen molecule \cite{Spalek2026a}.

\section*{Acknowledgments}
Many people have contributed to my understanding of the research in condensed matter theory. My Ph.D. supervisor, late Prof. Janusz Morkowski directed me towards the research on Hubbard model and the Mott phenomenon in the middle of 1970s. Prof. George Honig from Purdue showed me the intricacies of the metal insulator transition in pure and doped ${\rm V_2 O_3}$ and in ${\rm Ni S_{2-x}Se_x}$ from the experimental side. Profs.: Robert Gałązka, Tomasz Dietl, and Jacek Furdyna introduced me to the 
field of semimagnetic (diluted magnetic) semiconductors. Moreover, although most of the concepts in strongly 
correlated physics were of my own, Profs. Andrzej M. Oleś and Koung-An Chao
have been very helpful in developing them. Finally,
my own Ph.D. students at AGH University of Kraków, Purdue University, Warsaw University, and Jagiellonian University have been crucially important in my research over the last 30 years or so. Some of them are listed in the References, but  I am very grateful to all 24 of them. Finally, last but 
not least, the cooperation with Włodek Wójcik, Zbyszek Kąkol, Maciek Fidrysiak, Danuta Goc-Jagło, Piotrek Kuterba and Leszek Spałek was equally, if not more, important for my everyday work. I am indebted to P. Kuterba for technical help
in preparing this manuscript.

The research was supported by the Narodowe Centrum Nauki (NCN) Grant Nos. UMO-2021/41/B/ST3/04070 and 2023/49/B/ST3/03545.
\printbibliography

@Article{Kanamori1963,
  author    = {Kanamori, Junjiro},
  journal   = {Progress of Theoretical Physics},
  title     = {Electron Correlation and Ferromagnetism of Transition Metals},
  year      = {1963},
  issn      = {0033-068X},
  month     = sep,
  number    = {3},
  pages     = {275--289},
  volume    = {30},
  doi       = {10.1143/ptp.30.275},
  publisher = {Oxford University Press (OUP)},
}

@Article{Wysokinski2014,
  author    = {Wysokiński, Marcin M and Spałek, Jozef},
  journal   = {J. Phys.: Condens. Matter},
  title     = {Properties of an almost localized Fermi liquid in an applied magnetic field revisited: a statistically consistent Gutzwiller approach},
  year      = {2014},
  issn      = {1361-648X},
  month     = jan,
  number    = {5},
  pages     = {055601},
  volume    = {26},
  doi       = {10.1088/0953-8984/26/5/055601},
  publisher = {IOP Publishing},
}

@Article{Joerdens2008,
  author    = {Jördens, Robert and Strohmaier, Niels and Günter, Kenneth and Moritz, Henning and Esslinger, Tilman},
  journal   = {Nature},
  title     = {A Mott insulator of fermionic atoms in an optical lattice},
  year      = {2008},
  issn      = {1476-4687},
  month     = sep,
  number    = {7210},
  pages     = {204--207},
  volume    = {455},
  doi       = {10.1038/nature07244},
  publisher = {Springer Science and Business Media LLC},
}

@Book{Dirac2010,
  author    = {Dirac, P. A. M.},
  publisher = {Oxford University Press},
  title     = {The principles of quantum mechanics},
  year      = {2010},
  address   = {Oxford},
  edition   = {4. ed.},
  isbn      = {9780198520115},
  number    = {27},
  series    = {International series of monographs on physics},
  pagetotal = {314},
  ppn_gvk   = {1602991790},
}

@Article{Fock1957book,
  author    = {Fock, V. A.},
  title     = {Raboty po kvantovoi teorii polya},
  note      = {, pp. 25--51 (in Russian); (Izdatel'stvo Leningradskogo Universiteta, 1957);  for didactical exposition see A. L. Fetter and J. D. Walecka, \textit{Quantum Theory of Many-Particle Systems} (McGraw-Hill Book Co., 1971) pp.16-43}
}

@InCollection{Spalek2020,
  author    = {J{\'o}zef Spa{\l}ek},
  title     = {Mott Physics in Correlated Nanosystems: Localization-Delocalization Transition by the Exact Diagonalization Ab Initio Method},
  booktitle = {Topology, Entanglement, and Strong Correlations},
  editor    = {Eva Pavarini and Erik Koch},
  publisher = {Forschungszentrum Jülich Zentralbibliothek, Verlag},
  address   = {Jülich},
  year      = {2020},
  pages     = {7.1--7.38}
}

@Article{Spalek2007modifiedacta,
  note    = {For an early historical account see: J. Spałek, Acta Phys. Polon. A \textbf{111}, 409 (2007). For original detailed account see: J. Spa{\l}ek, Habilitation Thesis, Jagiellonian University, Krak\'ow (1981).}
}

@Article{Chao1978,
  author    = {Chao, K. A. and Spałek, J. and Oleś, A. M.},
  journal   = {Phys. Rev. B},
  title     = {Canonical perturbation expansion of the Hubbard model},
  year      = {1978},
  issn      = {0163-1829},
  month     = oct,
  number    = {7},
  pages     = {3453--3464},
  volume    = {18},
  doi       = {10.1103/physrevb.18.3453},
  publisher = {American Physical Society (APS)},
}

@Article{Spalek1988,
  author    = {Spałek, J.},
  journal   = {Phys. Rev. B},
  title     = {Effect of pair hopping and magnitude of intra-atomic interaction on exchange-mediated superconductivity},
  year      = {1988},
  issn      = {0163-1829},
  month     = jan,
  number    = {1},
  pages     = {533--536},
  volume    = {37},
  doi       = {10.1103/physrevb.37.533},
  publisher = {American Physical Society (APS)},
}

@Article{Yang1962,
  author    = {Yang, C. N.},
  journal   = {Rev. of Mod. Phys.},
  title     = {Concept of Off-Diagonal Long-Range Order and the Quantum Phases of Liquid He and of Superconductors},
  year      = {1962},
  issn      = {0034-6861},
  month     = oct,
  number    = {4},
  pages     = {694--704},
  volume    = {34},
  doi       = {10.1103/revmodphys.34.694},
  publisher = {American Physical Society (APS)},
}

@Article{Zegrodnik2017,
  author    = {Zegrodnik, Michał and Spałek, Józef},
  journal   = {Phys. Rev. B},
  title     = {Universal properties of high-temperature superconductors from real-space pairing: Role of correlated hopping and intersite Coulomb interaction within the $t$--$J$--$U$ model},
  year      = {2017},
  issn      = {2469-9969},
  month     = aug, 
  number    = {5},
  pages     = {054511},
  volume    = {96},
  doi       = {10.1103/physrevb.96.054511},
  publisher = {American Physical Society (APS)},
}

@Article{Dey2026,
author    = {Dey, Tushar and Fidrysiak, Maciej and Spałek, Józef},
  journal   = {},
  title     = {},
  year      = {2026},
  issn      = {},
  number    = {},
  pages     = {},
  volume    = {},
  doi       = {},
  publisher = {},
  note    = {, unpublished}
}

@Article{Kotliar1986,
  author    = {Kotliar, Gabriel and Ruckenstein, Andrei E.},
  journal   = {Phys. Rev. Lett.},
  title     = {New Functional Integral Approach to Strongly Correlated Fermi Systems: The Gutzwiller Approximation as a Saddle Point},
  year      = {1986},
  issn      = {0031-9007},
  month     = sep,
  number    = {11},
  pages     = {1362--1365},
  volume    = {57},
  doi       = {10.1103/physrevlett.57.1362},
  publisher = {American Physical Society (APS)},
}

@Article{Kaczmarczyk1008,
author    = {Kaczmarczyk, Jan and Jędrak, Jakub and Spałek, Józef},
journal = {arXiv: 1008.0021},
note = {(unpublished)}
}

@Article{Zegrodnik2018,
  author    = {Zegrodnik, Michał and Spałek, Józef},
  journal   = {Physical Review B},
  title     = {Incorporation of charge- and pair-density-wave states into the one--band model of d--wave superconductivity},
  year      = {2018},
  issn      = {2469-9969},
  month     = oct,
  number    = {15},
  pages     = {155144},
  volume    = {98},
  doi       = {10.1103/physrevb.98.155144},
  publisher = {American Physical Society (APS)},
}

@Misc{Bialo2020,
  author = {Bia{\l}o, I.},
  note   = {Ph.D. Thesis, AGH University of Science and Technology and TU Wien, 2020 (unpublished)}
}

@Article{Fidrysiak2018,
  note = {M. Fidrysiak, M. Zegrodnik, and J. Spałek, J. of Phys.: Condens. Matter \textbf{30}, 475602 (2018) and Refs. therein}
}

@Article{Fidrysiak2020,
  author    = {Fidrysiak, M. and Spałek, J.},
  journal   = {Phys. Rev. B},
  title     = {Robust spin and charge excitations throughout the high- Tc cuprate phase diagram from incipient Mottness},
  year      = {2020},
  issn      = {2469-9969},
  month     = jul,
  number    = {1},
  pages     = {014505},
  volume    = {102},
  doi       = {10.1103/physrevb.102.014505},
  publisher = {American Physical Society (APS)},
}

@Article{Fidrysiak2021,
  author    = {Fidrysiak, Maciej and Spałek, Józef},
  journal   = {Phys. Rev. B},
  title     = {Unified theory of spin and charge excitations in high--$T_c$ cuprate superconductors: A quantitative comparison with experiment and interpretation},
  year      = {2021},
  issn      = {2469-9969},
  month     = jul,
  number    = {2},
  pages     = {l020510},
  volume    = {104},
  doi       = {10.1103/physrevb.104.l020510},
  publisher = {American Physical Society (APS)},
}

@Article{Fidrysiak2021a,
  author    = {Fidrysiak, Maciej and Spałek, Józef},
  journal   = {Phys. Rev. B},
  title     = {Universal collective modes from strong electronic correlations: Modified $1/N_f$ theory with application to high--$T_c$ cuprates},
  year      = {2021},
  issn      = {2469-9969},
  month     = apr,
  number    = {16},
  pages     = {165111},
  volume    = {103},
  doi       = {10.1103/physrevb.103.165111},
  publisher = {American Physical Society (APS)},
}

@Article{Spalek1989,
  author    = {Spałek, J. and Gopalan, P.},
  journal   = {Journal de Physique},
  title     = {Exchange-mediated pairing: gap anisotropy and a narrow-band limit for hybridized electrons},
  year      = {1989},
  issn      = {0302-0738},
  number    = {18},
  pages     = {2869--2893},
  volume    = {50},
  doi       = {10.1051/jphys:0198900500180286900},
  publisher = {EDP Sciences},
}

@Article{KadzielawaMajor2014,
  author    = {Kądzielawa-Major, E. and Spałek, J.},
  journal   = {Acta Phys. Polon. A},
  title     = {Anderson-Kondo Lattice Hamiltonian from the Anderson-Lattice Model: A Modified Schrieffer-Wolff Transformation and the Effective Exchange Interactions},
  year      = {2014},
  issn      = {1898-794X},
  month     = oct,
  number    = {4A},
  pages     = {A-100-A-104},
  volume    = {126},
  doi       = {10.12693/aphyspola.126.a-100},
  publisher = {Institute of Physics, Polish Academy of Sciences},
}

@Article{Doradzinski1997,
  author    = {Doradziński, Roman and Spałek, Jozef},
  journal   = {Phys. Rev. B},
  title     = {Antiferromagnetic heavy-fermion and Kondo-insulating states with compensated magnetic moments},
  year      = {1997},
  issn      = {1095-3795},
  month     = dec,
  number    = {22},
  pages     = {R14239--R14242},
  volume    = {56},
  doi       = {10.1103/physrevb.56.r14239},
  publisher = {American Physical Society (APS)},
}

@Article{Doradzinski1998,
  author    = {Doradziński, Roman and Spałek, Jozef},
  journal   = {Phys. Rev. B},
  title     = {Mean-field magnetic phase diagram of the periodic Anderson model with the Kondo-compensated phases},
  year      = {1998},
  issn      = {1095-3795},
  month     = aug,
  number    = {6},
  pages     = {3293--3301},
  volume    = {58},
  doi       = {10.1103/physrevb.58.3293},
  publisher = {American Physical Society (APS)},
}

@Article{Howczak2012,
  author    = {Howczak, Olga and Spałek, Jozef},
  journal   = {J. Phys.: Condens. Matter},
  title     = {Anderson lattice with explicit Kondo coupling revisited: metamagnetism and the field-induced suppression of the heavy fermion state},
  year      = {2012},
  issn      = {1361-648X},
  month     = apr,
  number    = {20},
  pages     = {205602},
  volume    = {24},
  doi       = {10.1088/0953-8984/24/20/205602},
  publisher = {IOP Publishing},
}

@InProceedings{SpalekDoradzinski1998,
  author    = {Spa{\l}ek, J. and Doradzi{\'n}ski, R.},
  booktitle = {Magnetism and Electronic Correlations in Local-Moment Systems: Rare-Earth Elements and Compounds},
  editor    = {Donath, M. and Dowben, P. A. and Nolting, W.},
  publisher = {World Scientific, Singapore},
  pages = {387-405},
  year      = {1998}
}

@Article{Steglich1979,
  author    = {Steglich, F. and Aarts, J. and Bredl, C. D. and Lieke, W. and Meschede, D. and Franz, W. and Schäfer, H.},
  journal   = {Physical Review Letters},
  title     = {Superconductivity in the Presence of Strong Pauli Paramagnetism: $CeCu_2Si_2$},
  year      = {1979},
  issn      = {0031-9007},
  month     = dec,
  number    = {25},
  pages     = {1892--1896},
  volume    = {43},
  doi       = {10.1103/physrevlett.43.1892},
  publisher = {American Physical Society (APS)},
}

@Article{Andres1975,
  author    = {Andres, K. and Graebner, J. E. and Ott, H. R.},
  journal   = {Phys. Rev. Lett.},
  title     = {4f-Virtual-Bound-State Formation in $CeAl_3$ at Low Temperatures},
  year      = {1975},
  issn      = {0031-9007},
  month     = dec,
  number    = {26},
  pages     = {1779--1782},
  volume    = {35},
  doi       = {10.1103/physrevlett.35.1779},
  publisher = {American Physical Society (APS)},
}

@Article{Spalek1988a,
  author    = {Spałek, Józef},
  journal   = {Phys. Rev. B},
  title     = {Microscopic model of hybrid pairing: A common approach to heavy-fermion and high-$T_c$ superconductivity},
  year      = {1988},
  issn      = {0163-1829},
  month     = jul,
  number    = {1},
  pages     = {208--212},
  volume    = {38},
  doi       = {10.1103/physrevb.38.208},
  publisher = {American Physical Society (APS)},
}

@Article{Zegrodnik2019,
  author    = {Zegrodnik, M. and Biborski, A. and Fidrysiak, M. and Spałek, J.},
  journal   = {Phys. Rev. B},
  title     = {Superconductivity in the three-band model of cuprates: Variational wave function study and relation to the single-band case},
  year      = {2019},
  issn      = {2469-9969},
  month     = mar,
  number    = {10},
  pages     = {104511},
  volume    = {99},
  doi       = {10.1103/physrevb.99.104511},
  publisher = {American Physical Society (APS)},
}

@Article{Spalek2006a,
  author    = {Spałek, Jozef},
  journal   = {Physica B: Condensed Matter},
  title     = {Spin-split masses and a critical behavior of almost localized narrow-band and heavy-fermion systems},
  year      = {2006},
  issn      = {0921-4526},
  month     = may,
  pages     = {654--660},
  volume    = {378–380},
  doi       = {10.1016/j.physb.2006.01.418},
  publisher = {Elsevier BV},
}

@Article{Sheikin2003,
  author    = {Sheikin, I. and Gröger, A. and Raymond, S. and Jaccard, D. and Aoki, D. and Harima, H. and Flouquet, J.},
  journal   = {Phys. Rev. B},
  title     = {High magnetic field study of $CePd_2Si_2$},
  year      = {2003},
  issn      = {1095-3795},
  month     = mar,
  number    = {9},
  pages     = {094420},
  volume    = {67},
  doi       = {10.1103/physrevb.67.094420},
  publisher = {American Physical Society (APS)},
}

@Article{McCollam2005,
  author    = {McCollam, A. and Julian, S. R. and Rourke, P. M. C. and Aoki, D. and Flouquet, J.},
  journal   = {Phys. Rev. Lett.},
  title     = {Anomalous {de Haas–van Alphen} Oscillations in $CeCoIn_5$},
  year      = {2005},
  issn      = {1079-7114},
  month     = may,
  number    = {18},
  pages     = {186401},
  volume    = {94},
  doi       = {10.1103/physrevlett.94.186401},
  publisher = {American Physical Society (APS)},
}

@Article{Semeniuk2023,
  author    = {Semeniuk, Konstantin and Chang, Hui and Baglo, Jordan and Friedemann, Sven and Tozer, Stanley W. and Coniglio, William A. and Gamża, Monika B. and Reiss, Pascal and Alireza, Patricia and Leermakers, Inge and McCollam, Alix and Grockowiak, Audrey D. and Grosche, F. Malte},
  journal   = {Proceedings of the National Academy of Sciences},
  title     = {Truncated mass divergence in a Mott metal},
  year      = {2023},
  issn      = {1091-6490},
  month     = sep,
  number    = {38},
  volume    = {120},
  doi       = {10.1073/pnas.2301456120},
  publisher = {Proceedings of the National Academy of Sciences},
}

@Article{Citro1999,
  author    = {Citro, R. and Romano, A. and Spałek, J.},
  journal   = {Physica B: Condensed Matter},
  title     = {Kondo-lattice in an applied magnetic field: spin-split masses and metamagnetism},
  year      = {1999},
  issn      = {0921-4526},
  month     = jan,
  pages     = {213--214},
  volume    = {259–261},
  doi       = {10.1016/s0921-4526(98)00988-0},
  publisher = {Elsevier BV},
}

@Article{Kurleto2021,
  author    = {Kurleto, R. and Fidrysiak, M. and Nicolaï, L. and Minár, J. and Rosmus, M. and Walczak, Ł. and Tejeda, A. and Rault, J. E. and Bertran, F. and Kądzielawa, A. P. and Legut, D. and Gnida, D. and Kaczorowski, D. and Kissner, K. and Reinert, F. and Spałek, J. and Starowicz, P.},
  journal   = {Phys. Rev. B},
  title     = {Photoemission signature of momentum-dependent hybridization in $CeCoIn_5$},
  year      = {2021},
  issn      = {2469-9969},
  month     = sep,
  number    = {12},
  pages     = {125104},
  volume    = {104},
  doi       = {10.1103/physrevb.104.125104},
  publisher = {American Physical Society (APS)},
}

@Article{Spalek2026n,
year = {2026},
note = {J. Spałek, P. Kuterba, M. Wójcik, et al., in preparation.}
}

@Article{Pavarini2022n,
  note = {See, e.g., \textit{Dynamical Mean-Field Theory of Correlated Electrons}, edited by E. Pavarini et al., Schriften des Forschungszentrums J\"ulich, Vol.~12 (2022)}
}

@Article{Lieb1968,
  author    = {Lieb, Elliott H. and Wu, F. Y.},
  journal   = {Phys. Rev. Lett.},
  title     = {Absence of Mott Transition in an Exact Solution of the Short-Range, One-Band Model in One Dimension},
  year      = {1968},
  issn      = {0031-9007},
  month     = jun,
  number    = {25},
  pages     = {1445--1448},
  volume    = {20},
  doi       = {10.1103/physrevlett.20.1445},
  publisher = {American Physical Society (APS)},
}

@Article{Lieb2003,
  author    = {Lieb, Elliott H. and Wu, F.Y.},
  journal   = {Physica A: Statistical Mechanics and its Applications},
  title     = {The one-dimensional Hubbard model: a reminiscence},
  year      = {2003},
  issn      = {0378-4371},
  month     = apr,
  number    = {1–2},
  pages     = {1--27},
  volume    = {321},
  doi       = {10.1016/s0378-4371(02)01785-5},
  publisher = {Elsevier BV},
}

@Article{Kurzyk2007,
  author    = {Kurzyk, J. and Spałek, J. and Wójcik, W.},
  journal   = {Acta Phys. Polon. A},
  title     = {{Lieb-Wu} Solution, Gutzwiller-Wave-Function, and Gutzwiller-Ansatz Approximations with Adjustable Single-Particle Wave Function for the Hubbard Chain},
  year      = {2007},
  issn      = {1898-794X},
  month     = apr,
  number    = {4},
  pages     = {603--618},
  volume    = {111},
  doi       = {10.12693/aphyspola.111.603},
  publisher = {Institute of Physics, Polish Academy of Sciences},
}

@Article{Kurzyk2008,
  author    = {Kurzyk, J. and Wójcik, W. and Spałek, J.},
  journal   = {Eur. Phys. J. B},
  title     = {Extended Hubbard model with renormalized Wannier wave functions in the correlated state: beyond the parametrized models},
  year      = {2008},
  issn      = {1434-6036},
  month     = dec,
  number    = {3},
  pages     = {385--398},
  volume    = {66},
  doi       = {10.1140/epjb/e2008-00433-1},
  publisher = {Springer Science and Business Media LLC},
}

@Article{Spalek2010,
  author    = {Spałek, J. and Kurzyk, J. and Podsiadły, R. and Wójcik, W.},
  journal   = {Eur. Phys. J. B},
  title     = {Extended Hubbard model with the renormalized Wannier wave functions in the correlated state II: quantum critical scaling of the wave function near the Mott-Hubbard transition},
  year      = {2010},
  issn      = {1434-6036},
  month     = mar,
  number    = {1},
  pages     = {63--74},
  volume    = {74},
  doi       = {10.1140/epjb/e2010-00077-6},
  publisher = {Springer Science and Business Media LLC},
}

@Article{Hendzel2022,
  author    = {Hendzel, Maciej and Fidrysiak, Maciej and Spałek, Józef},
  journal   = {J. Phys. Chem. Lett.},
  title     = {Toward Complementary Characterization of the Chemical Bond},
  year      = {2022},
  issn      = {1948-7185},
  month     = oct,
  number    = {44},
  pages     = {10261--10266},
  volume    = {13},
  doi       = {10.1021/acs.jpclett.2c02544},
  publisher = {American Chemical Society (ACS)},
}

@InCollection{Broclawik2023,
  author    = {Broc{\l}awik, E. and Fidrysiak, M. and Hendzel, M. and Spa{\l}ek, J.},
  title     = {Interparticle correlations and chemical bonding from physical side: Covalency vs atomicity and ionicity},
  booktitle = {Polish Quantum Chemistry from Ko{\l}os to Now},
  publisher = {Elsevier},
  year      = {2023},
  pages     = {351--373},
  doi       = {10.1016/bs.aiq.2023.02.002}
}

@Article{Spalek2007a,
  author    = {Spałek, Jozef and Görlich, Edward M and Rycerz, Adam and Zahorbeński, Roman},
  journal   = {J. Phys.: Condens. Matter},
  title     = {The combined exact diagonalization–ab initio approach and its application to correlated electronic states and {Mott--Hubbard} localization in nanoscopic systems},
  year      = {2007},
  issn      = {1361-648X},
  month     = may,
  number    = {25},
  pages     = {255212},
  volume    = {19},
  doi       = {10.1088/0953-8984/19/25/255212},
  publisher = {IOP Publishing},
}

@Article{SpalekWojcik1995n,
  note = {For a review see: J. Spa{\l}ek and W. W{\'o}jcik, \textit{Almost Localized Fermions and Mott-Hubbard Transitions at Non-Zero Temperature}, in \textit{Spectroscopy of Mott Insulators and Correlated Metals}, Springer Series in Solid State Sciences, Vol.~19, pp.~41--65 (1995)}
}

@Article{Spalek2022,
  author    = {Spałek, J. and Fidrysiak, M. and Zegrodnik, M. and Biborski, A.},
  journal   = {Phys. Rep.},
  title     = {Superconductivity in high-$T_c$ and related strongly correlated systems from variational perspective: Beyond mean field theory},
  year      = {2022},
  issn      = {0370-1573},
  month     = may,
  pages     = {1--117},
  volume    = {959},
  doi       = {10.1016/j.physrep.2022.02.003},
  publisher = {Elsevier BV},
}

@Article{Hashimoto2014n,
  note = {For detailed experimental results review see: Hashimoto, M. and Vishik, I. M. and He, R.-H. and Devereaux, T. P. and Shen, Z.-X., Nature Physics, \textbf{10}, 483 (2014)}
}

@Misc{Howczak2012phd,
  author = {Howczak, O.},
  note   = {Ph.D. Thesis, Jagiellonian University, Kraków, 2012}
}

@Article{Hendzel2026,
note = {M. Hendzel, P. Kuterba, J. Spałek, Role of Kinetic Exchange and Coulomb
Interaction in Bonding of Hydrogen Molecular Systems and Excited States,
Acta Phys. Pol. B 57, 5-A18 (2026), this issue.}
}

@Article{Fidrysiak2026,
note = { M. Fidrysiak, Fermiology, Charge Transfer Energy, and Robust
Paramagnons in high-$T_c$ Cuprate Superconductors, Acta Phys. Pol. B 57,
5-A14 (2026), this issue.}
}

@Article{Spalek2026etal,
note = {J. Spałek et al., A Nonstandard Statistics for Strongly Correlated Systems:
Two Simple Examples, Acta Phys. Pol. B 57, 5-A16 (2026), this issue.}
}

@Article{Spalek2026a,
author    = {Spałek, Józef, and Hendzel, Maciej},
year = {2026},
note = {in preparation}
}

@Book{2013,
  publisher = {Springer Berlin Heidelberg},
  title     = {Strongly Correlated Systems: Numerical Methods},
  year      = {2013},
  isbn      = {9783642351068},
  editor    = {Avella, A. and Mancini, F.},
  doi       = {10.1007/978-3-642-35106-8},
  issn      = {0171-1873},
  journal   = {Springer Series in Solid-State Sciences},
}

@Article{Spalek1977,
  author    = {Spałek, J. and Oleś, A.M.},
  journal   = {Physica B+C},
  title     = {Ferromagnetism in narrow s-band with inclusion of intersite correlations},
  year      = {1977},
  issn      = {0378-4363},
  month     = jan,
  pages     = {375--377},
  volume    = {86–88},
  doi       = {10.1016/0378-4363(77)90352-7},
  publisher = {Elsevier BV},
}

@Article{Chao1977,
  author    = {Chao, K A and Spałek, J and Oleś, A M},
  journal   = {J. of Phys. C: Solid State Physics},
  title     = {Kinetic exchange interaction in a narrow S-band},
  year      = {1977},
  issn      = {0022-3719},
  month     = may,
  number    = {10},
  pages     = {L271--L276},
  volume    = {10},
  doi       = {10.1088/0022-3719/10/10/002},
  publisher = {IOP Publishing},
}

@Article{Spalek1980,
  author    = {Spałek, J and Chao, K A},
  journal   = {Journal of Physics C: Solid State Physics},
  title     = {Kinetic exchange interaction in a doubly degenerate narrow band and its application to $Fe_{1-x}Co_xS_2$ and $Co_{1-x} Ni_x S_2$},
  year      = {1980},
  issn      = {0022-3719},
  month     = oct,
  number    = {28},
  pages     = {5241--5251},
  volume    = {13},
  doi       = {10.1088/0022-3719/13/28/011},
  publisher = {IOP Publishing},
}

@Article{Spalek2017,
  author    = {Spałek, Józef and Zegrodnik, Michał and Kaczmarczyk, Jan},
  journal   = {Phys. Rev. B},
  title     = {Universal properties of high-temperature superconductors from real-space pairing: $t$--$J$--$U$ model and its quantitative comparison with experiment},
  year      = {2017},
  issn      = {2469-9969},
  month     = jan,
  number    = {2},
  pages     = {024506},
  volume    = {95},
  doi       = {10.1103/physrevb.95.024506},
  publisher = {American Physical Society (APS)},
}

@Article{Spalek1990a,
  author    = {Spałek, J. and Gopalan, P.},
  journal   = {Phys. Rev. Lett.},
  title     = {Almost-localized electrons in a magnetic field},
  year      = {1990},
  issn      = {0031-9007},
  month     = jun,
  number    = {23},
  pages     = {2823--2826},
  volume    = {64},
  doi       = {10.1103/physrevlett.64.2823},
  publisher = {American Physical Society (APS)},
}

@Article{Korbel1995,
  author    = {Korbel, P. and Spałek, J. and Wójcik, W. and Acquarone, M.},
  journal   = {Phys. Rev. B},
  title     = {Spin-split masses and metamagnetic behavior of almost-localized fermions},
  year      = {1995},
  issn      = {1095-3795},
  month     = jul,
  number    = {4},
  pages     = {R2213--R2216},
  volume    = {52},
  doi       = {10.1103/physrevb.52.r2213},
  publisher = {American Physical Society (APS)},
}

@Article{Spalek2006,
  author    = {Spałek, Józef},
  journal   = {Physica Status Solidi (b)},
  title     = {Magnetic properties of almost localized fermions revisited: spin dependent masses and quantum critical behavior},
  year      = {2006},
  issn      = {1521-3951},
  month     = jan,
  number    = {1},
  pages     = {78--88},
  volume    = {243},
  doi       = {10.1002/pssb.200562526},
  publisher = {Wiley},
}

@Article{Shishido2018,
  author    = {Shishido, Hiroaki and Yamada, Shogo and Sugii, Kaori and Shimozawa, Masaaki and Yanase, Youichi and Yamashita, Minoru},
  journal   = {Phys. Rev. Lett.},
  title     = {Anomalous Change in the de {Haas–-van Alphen} Oscillations of $CeCoIn_5$ at Ultralow Temperatures},
  year      = {2018},
  issn      = {1079-7114},
  month     = apr,
  number    = {17},
  pages     = {177201},
  volume    = {120},
  doi       = {10.1103/physrevlett.120.177201},
  publisher = {American Physical Society (APS)},
}

@Book{Mott1997n,
  note = {For a review see, e.g., N. F. Mott, \textit{Metal-Insulator Transitions}, 2 ed., Taylor \& Francis},
  address = {London},
  year = {1991}
}

@Article{Anderson1959,
  author    = {Anderson, P. W.},
  journal   = {Phys. Rev.},
  title     = {New Approach to the Theory of Superexchange Interactions},
  year      = {1959},
  issn      = {0031-899X},
  month     = jul,
  number    = {1},
  pages     = {2--13},
  volume    = {115},
  doi       = {10.1103/physrev.115.2},
  publisher = {American Physical Society (APS)},
}

@InBook{Anderson1963,
  author    = {Anderson, Philip W.},
  pages     = {99--214},
  publisher = {Elsevier},
  title     = {Theory of Magnetic Exchange Interactions: Exchange in Insulators and Semiconductors},
  year      = {1963},
  volume = {14},
  editor = {Seitz, F. and Turnbull, S.},
  isbn      = {9780126077148},
  booktitle = {Solid State Physics},
  doi       = {10.1016/s0081-1947(08)60260-x},
  issn      = {0081-1947},
}

@Article{Hubbard1964,
  author    = {Hubbard, J.},
  journal   = {Proc. Roy. Soc. (London) S.},
  title     = {Electron correlations in narrow energy bands III. An improved solution},
  year      = {1964},
  issn      = {2053-9169},
  month     = sep,
  number    = {1386},
  pages     = {401--419},
  volume    = {281},
  doi       = {10.1098/rspa.1964.0190},
  publisher = {The Royal Society},
}

@Article{Gutzwiller1965,
  author    = {Gutzwiller, Martin C.},
  journal   = {Phys. Rev.},
  title     = {Correlation of Electrons in a Narrow Band},
  year      = {1965},
  issn      = {0031-899X},
  month     = mar,
  number    = {6A},
  pages     = {A1726--A1735},
  volume    = {137},
  doi       = {10.1103/physrev.137.a1726},
  publisher = {American Physical Society (APS)},
}

@Article{Brinkman1970,
  author    = {Brinkman, W. F. and Rice, T. M.},
  journal   = {Phys. Rev. B},
  title     = {Application of Gutzwiller’s Variational Method to the Metal-Insulator Transition},
  year      = {1970},
  issn      = {0556-2805},
  month     = nov,
  number    = {10},
  pages     = {4302--4304},
  volume    = {2},
  doi       = {10.1103/physrevb.2.4302},
  publisher = {American Physical Society (APS)},
}

@Article{Spalek1987,
  author    = {Spałek, J. and Datta, A. and Honig, J.M.},
  journal   = {Phys. Rev. Lett.},
  title     = {Discontinuous metal-insulator transitions and Fermi-liquid behavior of correlated electrons},
  year      = {1987},
  issn      = {0031-9007},
  month     = aug,
  number    = {6},
  pages     = {728--731},
  volume    = {59},
  doi       = {10.1103/physrevlett.59.728},
  publisher = {American Physical Society (APS)},
}

@Article{Spaek1989,
  author    = {Spałek, J. and Kokowski, M. and Honig, J. M.},
  journal   = {Phys. Rev. B},
  title     = {Low-temperature properties of an almost-localized Fermi liquid},
  year      = {1989},
  issn      = {0163-1829},
  month     = mar,
  number    = {7},
  pages     = {4175--4185},
  volume    = {39},
  doi       = {10.1103/physrevb.39.4175},
  publisher = {American Physical Society (APS)},
}

@Article{Spalek1990,
  author    = {Spałek, Jozef},
  journal   = {J. of Sol. St. Chem.},
  title     = {Fermi liquid behavior and the metal-insulator transition of almost localized electrons: A brief theoretical review and an application to $V_2O_3$ system},
  year      = {1990},
  issn      = {0022-4596},
  month     = sep,
  number    = {1},
  pages     = {70--93},
  volume    = {88},
  doi       = {10.1016/0022-4596(90)90206-d},
  publisher = {Elsevier BV},
}

@Article{Spalek2007,
  author    = {Spa{\l}ek, Jozef},
  title     = {{t-J} model then and now: A personal perspective from the pioneering times},
  year      = {2007},
  journal = {Acta Phys. Polon. A},
  volume        = {111},
  pages         = {409--424},
  copyright = {Assumed arXiv.org perpetual, non-exclusive license to distribute this article for submissions made before January 2004},
  publisher = {arXiv},
}

@Article{Spalek2012,
  author    = {Spałek, J.},
  journal   = {Acta Phys. Polon. A},
  title     = {Theory of Unconventional Superconductivity in Strongly Correlated Systems: Real Space Pairing and Statistically Consistent Mean-Field Theory - in Perspective},
  year      = {2012},
  issn      = {1898-794X},
  month     = apr,
  number    = {4},
  pages     = {764--784},
  volume    = {121},
  doi       = {10.12693/aphyspola.121.764},
  publisher = {Institute of Physics, Polish Academy of Sciences},
}

@Article{Spalek2023,
  author    = {Spałek, J.},
  journal   = {Acta Phys. Polon. A},
  title     = {Brief Perspective of High-Temperature Superconductivity in the Cuprates: Strong Correlations Combined with Superexchange Match Experiment},
  year      = {2023},
  issn      = {0587-4246},
  month     = feb,
  number    = {2},
  pages     = {169--179},
  volume    = {143},
  doi       = {10.12693/aphyspola.143.169},
  publisher = {Institute of Physics, Polish Academy of Sciences},
}

@Article{Jones2015,
  note = {See, e.g., R.O. Jones, "Density functional theory: Its origins, rise to prominence, and future", Rev. Mod. Phys. \textbf{87}, 897 (2015).},
}
\end{document}